# The Periodic Table of Elementary Particles and the Composition of Hadrons

Ding-Yu Chung

All leptons, quarks, and gauge bosons can be placed in the periodic table of elementary particles. As the periodic table of elements derived from atomic orbital, the periodic table of elementary particles is derived from the two sets of seven orbitals: principal dimensional orbital and auxiliary dimensional orbital. (Seven orbitals come indirectly from the seven extra dimensions in eleven-dimensional space-time.) Principal dimensional orbital derived from varying space-time dimension, varying speed of light, and varying supersymmetry explains gauge bosons and low-mass leptons. Auxiliary dimensional orbital derived from principal dimensional orbital accounts for high-mass leptons and individual quarks. For hadrons as the composites of individual quarks, hadronic dimensional orbital derived from auxiliary dimensional orbital is responsible. These three sets of seven orbitals explain all elementary particles and hadrons. QCD, essentially, describes the different occupations of quarks in the three sets of seven orbitals at different temperatures. The periodic table of elementary particles and the compositions of hadrons relate to the Barut lepton mass formula, the Polazzi mass formula for stable hadrons, and the MacGregor-Akers constituent quark model. The calculated masses for elementary particles and hadrons are in good agreement with the observed masses. For examples, the calculated masses for the top quark, neutron, and pion are 176.5 GeV, 939.54MeV, and 135.01MeV in excellent agreement with the observed masses, 174.3 ± 5.1GeV, 939.57 MeV, and 134.98 MeV, respectively.

## 1. *Introduction*

The accurate calculation for the masses of elementary particles first appeared in the quantized lepton mass formula by A.O. Barut [1].

$$M_n = M_e + \frac{3M_e}{2\alpha} \sum_{n=0}^{n} n^4$$

where n = 0, 1, and 2 for electron, muon, and τ, respectively. The calculated masses 105.5 and 1786 MeV for muon and τ, respectively, in good agreement with the observed masses, 105.7 and 1777 MeV. The quantized lepton mass formula indicates the orbital structure. Using this orbital structure, the periodic table of elementary particles accounts for all leptons, quarks, and gauge bosons [2].

P. Polazzi finds that the cube root of the masses for stable hadrons as the function of order of these stable hadrons is a straight line, which also occurs for stable atoms [3]. This is an indication of orbital structure for hadrons. Both the periodic table of elementary particles and the Polazzi mass formula for stable hadrons involve constituent masses as in MacGregor- Akers constituent quark model [4] [5] that can calculate the masses of hadrons accurately.

This paper posits that the periodic table of elementary particles and the composition of hadrons are derived from the three sets of seven orbitals: principal dimensional orbital, auxiliary dimensional orbital, and hadronic dimensional orbital. (Seven orbitals come indirectly from the seven extra dimensions in eleven-dimensional space-time.) Sections 2 is about the origin of principal dimensional orbital from space-time dimension, varying speed of light, and varying supersymmetry. Section 3 is about principal dimensional orbital for gauge bosons. Section 4 is about the combination of principal dimensional orbital and auxiliary dimensional orbital for the periodic table of elementary particles. Section 5 is about the combination of auxiliary orbital and hadronic orbital for the compositions of hadrons. Section 6 deals with QCD and the three sets of seven orbitals.

## 2. *The varying dimension number transformation: the origin of principal dimensional orbital*

The constancy of the speed of light is the pillar of special relativity. The constancy of the speed of light takes place in the four dimensional space-time whose space-time dimension number (four) is constant. In the varying speed of light (VSL) model [6] of cosmology by Albrecht, Magueijo, and Barrow, the speed of light varies in time. The time dependent speed of light varies as some power of the expansion scale factor, a, in such way that

$$c(t)=c_0\, a^n \tag{1}$$

where $c_0 > 0$ and n are constant. The increase of speed of light is continuous.

This paper posits quantized varying speed of light (QVSL), where the speed of light is invariant in a constant space-time dimension number, and the speed of light varies with varying space-time dimension number from 4 to 11. In QVSL, the speed of light is quantized by varying space-time dimension number.

$$c_D = c\,/\,\alpha^{D-4}, \tag{2}$$

where c is the observed speed of light in the 4D space-time, $c_D$ is the quantized varying speed of light in space-time dimension number, D, from 4 to 11, and α is the fine structure constant. Each dimensional space-time has a specific speed of light. The speed of light increases with increasing space-time dimension number, D. In the VDN model of cosmology [7], the universe starts with the pre-expanding universe that has the speed of light in 11D space-time.

In special relativity, $E = M_0\, c^2$ modified by eq. (2) is expressed as

$$E = M_0 (c^2\,/\,\alpha^{2(D-4)}) \tag{3a}$$

$$= (M_0\,/\,\alpha^{2(d-4)})\, c^2 \tag{3b}$$

Eq. (3a) means that a particle in the D dimensional space-time can have superluminal speed, $c\,/\,\alpha^{D-4}$, that is higher than the observed speed of light, and has rest mass, $M_0$. Eq.

(3b) means that the same particle in the 4D space-time with the observed speed of light acquires $M_0/\alpha^{2(d-4)}$ as the rest mass where d = D. D in eq. (3a) is space-time dimension number defining the varying speed of light. In eq. (3b), d from 4 to 11 is "mass dimension number" defining varying mass. For example, for D = 11, eq. (3a) shows a superluminal particle in eleven-dimensional space-time, while eq. (3b) shows that the speed of light of the same particle is the observed speed of light with the 4D space-time, and the mass dimension is eleven. In other words, 11D space-time can transform into 4D space-time with 11d mass dimension. QVSL in terms of varying space-time dimension number, D, brings about varying mass in terms of varying mass dimension number, d.

The QVSL transformation transforms space-time dimension number and mass dimension number. In the QVSL transformation, the decrease in the speed of light leads to the decrease in space-time dimension number and the increase of mass in terms of increasing mass dimension number from 4 to 11.

$$c_D = c_{D-n} / \alpha^{2n}, \tag{4a}$$

$$M_{0,D,d} = M_{0,D-n,\ d+n} \alpha^{2n}, \tag{4b}$$

$$D,\ d \xrightarrow{QVSL} (D \mp n),\ \ (d \pm n) \tag{4c}$$

where D is space-time dimension number from 4 to 11 and d is mass dimension number from 4 to 11. For example, the QVSL transformation transforms a particle with 11D4d to a particle with 4D11d. In terms of rest mass, 11D space-time has 4d with the lowest rest mass, and 4D space-time has 11d with the highest rest mass.

The QVSL transformation is an alternate to the Higgs mechanism to gain rest mass. In the QVSL, the speed of light is constant in a specific space-time dimension number, such as 4 for our four-dimensional space-time. In different space-time dimension numbers (from 4 to 11), speeds of light are different. In our four-dimensional space-time, the speed of light is the lowest, so according to special relativity ($E = m_0c^2$), with constant energy, the rest mass in our four-dimensional space-time is the highest. Thus, instead of absorbing the Higgs boson to gain rest mass, a particle can gain rest mass by decreasing the speed of light and space-time dimension number. The QVSL transformation also gain a new quantum number, "mass dimension number" from 4 to 11 to explain the hierarchical masses of elementary particles. Since the Higgs bosons have not been found experimentally, the QVSL transformation to gain rest mass is a good alternate. In terms of vacuum energy, the four-dimensional space-time has zero vacuum energy with the highest rest mass, while D > 4 have non-zero vacuum energy with lower rest mass than 4D.

Since the speed of light for > 4D particle is greater than the speed of light for 4D particle, the observation of > 4D particles by 4D particles violates casualty. Thus, > 4D particles are hidden particles with respect to 4D particles. Such hidden particles form the base for phantom energy [9].

In the normal supersymmetry transformation, the repeated application of the fermion-boson transformation transforms a boson (or fermion) from one point to the same boson (or fermion) at another point at the same mass. In the "varying supersymmetry transformation", the repeated application of the fermion-boson transformation transforms a boson from one point to the boson at another point at different mass dimension number in the same space-time number. The repeated varying supersymmetry transformation transforms boson $B_d$ into fermion $F_d$ and from fermion $F_d$ to boson $B_{d-1}$ is expressed as

$$M_{d,F} = M_{d,B}\ \alpha_{d,B}, \tag{5a}$$

$$M_{d-1,B} = M_{d,F}\ \alpha_{d,F}, \tag{5b}$$

where $M_{d,\ B}$ and $M_{d,\ F}$ are the masses for a boson and a fermion, respectively, d is mass dimension number, and $\alpha_{d,\ B}$ or $\alpha_{d,\ F}$ is the fine structure constant, which is the ratio between the masses of a boson and its fermionic partner. Assuming $\alpha_{d,\ B} = \alpha_{d,\ F}$, the relation between the bosons in the adjacent dimensions, then, can be expressed as

$$M_{d-1,B} = M_{d,B}\ \alpha_d^2, \tag{5c}$$

Eq. 5 shows that it is possible to describe mass dimensions > 4 in terms of

$$F_5 B_5\ F_6\ B_6\ F_7\ B_7\ F_8\ B_8\ F_9\ B_9\ F_{10}\ B_{10}\ F_{11}\ B_{11} \quad , \tag{6}$$

where the energy of $B_{11}$ is Planck energy. Each mass dimension between 4d and 11d consists of a boson and a fermion. Eq. 5 shows a stepwise transformation that transforms a particle with d mass dimension to d ± 1 mass dimension. The transformation from higher dimensional particle to adjacent lower dimensional particle is the fractionalization of a higher dimensional particle to many lower dimensional particle in such way the number of lower dimensional particles = $n_{d-1} = n_d / \alpha^2$. The transformation from lower dimensional particles to higher dimensional particle is condensation. Both the fractionalization and the condensation are stepwise. For example, a particle with 4D (space-time) 10d (mass dimension) can transform stepwise into 4D9d particles. Since supersymmetry transformation involves translation, this stepwise varying supersymmetry transformation leads to translational fractionalization and translational condensation, resulting in expansion and contraction.

Another type of the varying supersymmetry transformation is not stepwise. It is the leaping varying supersymmetry transformation that transforms a particle with d mass dimension to any d ± n mass dimension. The transformation involves the fission-fusion of particle. The transformation from d to d – n involves the fission of a particle with d mass dimension into two parts: the core particle with d – n dimension and the dimensional orbitals that are separable from the core particle. The sum of the number of mass dimensions for a particle and the number of dimensional orbitals is equal to 11 for all particles with mass dimensions. Therefore,

$$F_d = F_{d-n} + (11 - d + n)\, DO's\ , \tag{7}$$

where 11- d + n is the number of dimensional orbitals (DO's) for $F_{d-n}$. For example, the fission of 4D9d particle produces 4D4d particle that has d = 4 core particle and 7 separable dimensional orbitals in the form of $B_5F_5B_6F_6B_7F_7B_8F_8B_9F_9B_{10}F_{10}B_{11}$. Since the fission process is not stepwise from higher mass dimension to lower mass dimension, it is possible to have simultaneous fission. For example, 4D9d particles can simultaneously transform into 4D8d, 4D7d, 4D6d, 4D5d, and 4D4d particles, which have 3, 4, 5, 6, and 7 separable dimensional orbitals, respectively. Therefore, varying supersymmetry transformation can be stepwise or leaping. Stepwise supersymmetry transformation is translational fractionalization and condensation, resulting in stepwise expansion and contraction. Leaping supersymmetry transformation is not translational, and it is fission and fusion, resulting possibly in simultaneous formation of different particles with separable dimensional orbitals.

In summary, the QVSL transformation transform space-time dimension number and mass dimension number. The varying supersymmetry transforms varying mass dimension number in the same space-time number as follows (D = space-time dimension number and d = mass dimension number).

$$D,\ d \xrightarrow{QVSL} (D \mp n),\ \ (d \pm n)$$

$$D,\ d \xrightarrow{stepwise\ or\ leaping\ \ varying\ \ supersymmetry} D,\ (d \pm 1)\ or\ D, (d \pm n)$$

Before the inflation, the universe is made of superstrings as 10D4d with another dimension for gravity. 10D4d superstring transforms through the QVSL transformation quickly into 4D10d particles, which then transforms and fractionalizes quickly through varying supersymmetry transformation into 4D9d, resulting in inflationary expansion. The inflationary expansion occurs between the energy for 4D10d = $E_{Planck}\,\alpha^2$ = 6 x $10^{14}$ GeV and the energy for 4D9d = $E_{10}\ \alpha^2$ = 3 x $10^{10}$ GeV. At the end of the inflationary expansion, all 4D9d particles undergo the simultaneous fission to generate equally by mass and number into 4D9d, 4D8d, 4D7d, 4D6d, 4D5d, and 4D4d particles. Baryonic matter is 4D4d, while dark matter consists of the other five types of particles. The mass ratio of dark matter to baryonic matter is 5 to 1 in agreement with the observation [8] that shows that the universe consists of 25% dark matter, 5% baryonic matter, and 70% dark energy. Afterward, quantum fluctuation and thermal expansion (the big bang) take place. In summary, the process is as follows.

$$10D4d \xrightarrow{QVSL\ transformation} 4D10d \xrightarrow{stepwise\ fractionalization,\ inflation}$$

$$4D9d \xrightarrow{simultaneous\ \ fission} 4D9d + 4D8d + 4D7d + 4D6d + 4D5d + 4D4d + radiation$$

$$\rightarrow quantum\ \ fluctuation + thermal\ cosmic\ expansion$$

The mechanism for the fission into core particle and dimensional orbital requires "detachment space" that detaches core particle and dimensional orbital. The physical source of detachment space comes from the empty space left behind in the formation of cosmic radiation derived from the annihilation (detachment) of matter [7]. The absorption of detachment space in the gap between core particle and dimensional orbital initiates the fission. However, the gap between core particle and dimensional orbital is not purely detachment space that creates permanent detachment. The space between core particle and dimensional orbital is "hybrid space" from combining attachment space (the space for core particle) and detachment space. Hybrid space has neither complete attachment nor complete detachment. Hybrid space can be described by the uncertainty principle in quantum mechanics as

$$\Delta x \Delta p \geq h/4\pi , \tag{8}$$

where x is position and p is momentum. For the uncertainty principle in quantum mechanics, $\Delta$x is the uncertainty in position and $\Delta$p is the uncertainty in momentum during the measurement of the position and momentum of a particle. For hybrid space, $\Delta$x is distance of the gap between core particle and dimensional orbital, and $\Delta$p is the momentum in the gap by the contact between dimensional orbital and core particle. $\Delta x\Delta p \geq h/4\pi$ shows that neither $\Delta$x nor $\Delta$p can be zero. Thus, there cannot be complete attachment where $\Delta x = 0$, and there cannot be complete detachment where $\Delta p = 0$. For hybrid space, the uncertainty principle without precise position and momentum becomes the gap principle without complete attachment and detachment. Furthermore, since the gap is microscopic and coherent in the microscopic and coherent system of core particle and its dimensional orbital, hybrid space is microscopic and coherent. Thus, hybrid space for all particles is the space for quantum mechanics.

4D4d particle (baryonic matter) has seven dimensional orbitals from d = 5 to d = 11, denoted as principal dimensional orbital that is the first set of the three sets of seven orbitals.

## *3. Gauge bosons: principal dimensional orbital*

For 4D4d particle, the lowest energy dimensional orbital in principal dimensional orbital is $E_{5,B}$. As Eq. 5b, the lowest energy for the Coulombic field is

$$\begin{aligned} E_{5,B} &= \alpha M_{6,F} \\ &= \alpha M_e , \end{aligned} \tag{9}$$

where $M_e$ is the rest energy of electron, and $\alpha = \alpha_e$ , the fine structure constant for the magnetic field. The bosons generated are called "principal dimensional bosons" or "$B_d$". From Eq. 5 with only $\alpha_e$, the mass of electron, the mass of $Z^0$, and the number of principal dimensional orbital, the masses of $B_d$ as the gauge boson can be calculated as shown in Table 1.

**Table 1.** The Energies of the Principal dimensional bosons (Gauge Bosons)
$B_d$ = principal dimensional boson, $\alpha = \alpha_e$

| $B_d$ | $E_d$ | GeV | Gauge Boson | Interaction |
|---|---|---|---|---|
| $B_5$ | $M_e\,\alpha$ | $3.7\text{x}10^{-6}$ | A | electromagnetic |
| $B_6$ | $M_e/\alpha$ | $7\text{x}10^{-2}$ | $\pi_{1/2}$ | strong |
| $B_7$ | $E_6/\alpha_w^2 \cos\theta_w$ | 91.177 | $Z_L^0$ | weak (left) |
| $B_8$ | $E_7/\alpha^2$ | $1.7\text{x}10^6$ | $X_R$ | CP (right) nonconservation |
| $B_9$ | $E_8/\alpha^2$ | $3.2\text{x}10^{10}$ | $X_L$ | CP (left) nonconservation |
| $B_{10}$ | $E_9/\alpha^2$ | $6.0\text{x}10^{14}$ | $Z_R^0$ | weak (right) |
| $B_{11}$ | $E_{10}/\alpha^2$ | $1.1\text{x}10^{19}$ | | Planck energy |

In Table 1, $\alpha_w$ is not same as $\alpha$ of the rest, because there is symmetry group mixing between $B_5$ and $B_7$ as the symmetry mixing in the standard theory of the electroweak interaction, and $\sin\theta_w$ is not equal to 1. As shown latter, $B_5$, $B_6$, $B_7$, $B_8$, $B_9$, and $B_{10}$ are A (massless photon), $\pi_{1/2}$, $Z_L^0$, $X_R$, $X_L$, and $Z_R^0$, respectively, responsible for the electromagnetic field, the strong interaction, the weak (left handed) interaction, the CP (right handed) nonconservation, the CP (left handed) nonconservation, and the P (right handed) nonconservation, respectively. The calculated value for $\theta_w$ is $29.69^0$ in good agreement with $28.7^0$ for the observed value of $\theta_w$ [10]. The calculated energy for $B_{11}$ is $1.1\text{x}10^{19}$ GeV in good agreement with the Planck mass, $1.2\text{x}10^{19}$ GeV. As shown later, the calculated masses of all gauge bosons are also in good agreement with the observed values. Most importantly, the calculation shows that exactly seven dimensional orbitals are needed for all fundamental interactions.

All six non-gravitational principal dimensional bosons are represented by the internal symmetry groups, consisting of two sets of symmetry groups, U(1), U(1), and SU(2) with the left-right symmetry. Each $B_d$ is represented by an internal symmetry group as follows.

$B_5$: U(1), $B_6$: U(1), $B_7$: $SU(2)_L$, $B_8$: $U(1)_R$, $B_9$: $U(1)_L$, $B_{10}$: $SU(2)_R$

The additional symmetry group, U(1) X $SU(2)_L$, is formed by the "mixing" of U(1) in $B_5$ and $SU(2)_L$ in $B_7$. This mixing is same as in the standard theory of the electroweak interaction.

As in the standard theory to the electroweak interaction, the boson mixing of U(1) and $SU(2)_L$ is to create electric charge and to generate the bosons for leptons and quarks by combining isospin and hypercharge. The hypercharges for both $e^+$ and $\nu$ are 1, while for both u and d quarks, they are 1/3. The electric charges for $e^+$ and $\nu$ are 1 and 0, respectively, while for u and d quarks, they are 2/3 and -1/3, respectively.

The gauge boson for the strong interaction is $B_6$, which has 70 MeV, about one half of the mass of pion. Section 6 will deal with the strong interaction.

The principal dimensional boson, $B_8$, is a CP violating boson, because $B_8$ is assumed to have the CP-violating $U(1)_R$ symmetry. The ratio of the force constants between the CP-invariant $W_L$ in $B_8$ and the CP-violating $X_R$ in $B_8$ is

$$\frac{G_8}{G_7} = \frac{\alpha\ E_7^2\ \cos^2\Theta_W}{\alpha_W\ E_8^2} = 5.3\ X\ 10^{-10}\ , \tag{10}$$

which is in the same order as the ratio of the force constants between the CP-invariant weak interaction and the CP-violating interaction with $|\Delta S| = 2$.

The principal dimensional boson, $B_9$ ($X_L$), has the CP-violating $U(1)_L$ symmetry. $B_9$ generates matter. The ratio of force constants between $X_R$ with CP conservation and $X_L$ with CP-nonconservation is

$$\frac{G_9}{G_8} = \frac{\alpha E_8^2}{\alpha E_9^2} = 2.8\ X\ 10^{-9}\ , \tag{11}$$

which is the ratio of the numbers between matter (dark and baryonic) and photons in the universe. During the simultaneous fission into dark matter and baryonic matter, the number of baryonic matter particles is 1/6 of total matter particles. Hence, the ratio of the numbers between baryonic matter and photons is about $4.7 \times 10^{-10}$, which is close to the ratio (around $5 \times 10^{-10}$) obtained by the big bang nucleosynthesis.

## *4. The periodic table of elementary particles: principal dimensional and auxiliary dimensional orbitals*

Auxiliary dimensional orbital [7] is derived from principal dimensional orbital. It is for high-mass leptons and individual quarks. Auxiliary dimensional orbital is the second set of the three sets of seven orbitals. The combination of dimensional auxiliary dimensional orbitals constitutes the periodic table for elementary particles as shown in Fig. 1 and Table 1.

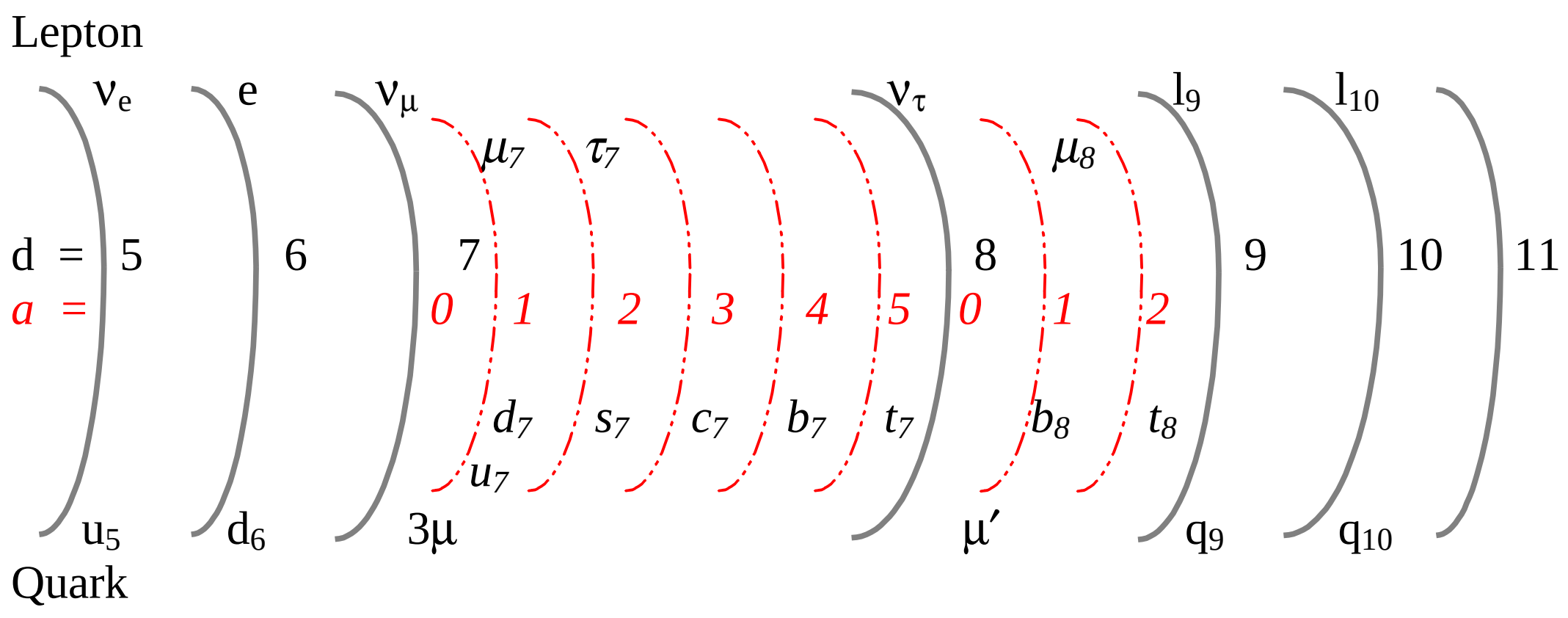

**Fig. 1:** leptons and quarks in the principal and auxiliary dimensional orbitals d = principal dimensional orbital (solid line) number, a = auxiliary dimensional orbital (dot line) number

**Table 2.** The Periodic Table of Elementary Particles
d = principal dimensional orbital number, a = auxiliary dimensional orbital number

| d | a = 0 | 1 | 2 | a = 0 | 1 | 2 | 3 | 4 | 5 | |
|---|---|---|---|---|---|---|---|---|---|---|
| | Lepton | | | Quark | | | | | | Boson |
| 5 | $l_5 = \nu_e$ | | | $q_5 = u = 3\nu_e$ | | | | | | $B_5 = A$ |
| 6 | $l_6 = e$ | | | $q_6 = d = 3e$ | | | | | | $B_6 = \pi_{1/2}$ |
| 7 | $l_7 = \nu_\mu$ | $\mu_7$ | $\tau_7$ | $q_7 = 3\mu$ | $u_7/d_7$ | $s_7$ | $c_7$ | $b_7$ | $t_7$ | $B_7 = Z_L^0$ |
| 8 | $l_8 = \nu_\tau$ | $\mu_8$ | | $q_8 = \mu'$ | $b_8$ | $t_8$ | | | | $B_8 = X_R$ |
| 9 | $l_9$ | | | $q_9$ | | | | | | $B_9 = X_L$ |
| 10 | | | | | | | | | | $B_{10} = Z_R^0$ |
| 11 | | | | | | | | | | $B_{11}$ |

"d" is the principal dimensional orbital number, and a is the auxiliary dimensional orbital number. (Note that $F_d$ has lower energy than $B_d$.)

There are two types of fermions in the periodic table of elementary particles: low-mass leptons and high-mass leptons and quarks. Low-mass leptons include $\nu_e$, e, $\nu_\mu$ , and $\nu_\tau$, which are in principal dimensional orbital, not in auxiliary dimensional orbital. $l_d$ is denoted as lepton with principal dimension number, d. $l_5$, $l_6$, $l_7$, and $l_8$ are $\nu_e$, e, $\nu_\mu$ , and $\nu_\tau$, respectively. All neutrinos have zero mass because of chiral symmetry (permanent chiral symmetry).

High-mass leptons and quarks include μ, τ, u, d, s, c, b, and t, which are the combinations of both principal dimensional fermions and auxiliary dimensional fermions. Each fermion can be defined by principal dimensional orbital numbers (d's) and auxiliary dimensional orbital numbers (a's) as $d_a$ in Table 3. For examples, e is $6_0$ that means it has d (principal dimensional orbital number) = 6 and a (auxiliary dimensional orbital number) = 0, so e is a principal dimensional fermion.

High-mass leptons, $\mu$ and $\tau$, are the combinations of principal dimensional fermions, e and $\nu_\mu$,, and auxiliary dimensional fermions. For example, $\mu$ is the combination of e, $\nu_\mu$,, and $\mu_7$, which is $7_1$ that has d = 7 and a = 1 .

Quarks are the combination of principal dimensional quarks ($q_d$) and auxiliary dimensional quarks. The principal dimensional fermion for quark is derived from principal dimensional lepton. To generate a principal dimensional quark in principal dimensional orbital from a lepton in the same principal dimensional orbital is to add the lepton to the boson from the combined lepton-antilepton. Thus, the mass of the quark is three times of the mass of the corresponding lepton in the same dimension. The equation for the mass of principal dimensional fermion for quark is

$$M_{q_d} = 3M_{l_d} \tag{12}$$

For principal dimensional quarks, $q_5$ ($5_0$) and $q_6$ ($6_0$) are $3\nu_e$ and 3e, respectively. Since $l_7$ is massless $\nu_\mu$ , $\nu_\mu$ is replaced by $\mu$, and $q_7$ is $3\mu$. Quarks are the combinations of principal dimensional quarks, $q_d$, and auxiliary dimensional quarks. For example, s quark is the combination of $q_6$ (3e), $q_7$ ($3\mu$) and $s_7$ (auxiliary dimensional quark = $7_2$).

Each fermion can be defined by principal dimensional orbital numbers (d's) and auxiliary dimensional orbital numbers (a's). All leptons and quarks with d's, a's and the calculated masses are listed in Table 3.

**Table 3.** The Compositions and the Constituent Masses of Leptons and Quarks
d = principal dimensional orbital number and a = auxiliary dimensional orbital number

| | da | Composition | Calculated Mass |
|---|---|---|---|
| Leptons | da for leptons | | |
| $\nu_e$ | $5_0$ | $\nu_e$ | 0 |
| e | $6_0$ | e | 0.51 MeV (given) |
| $\nu_\mu$ | $7_0$ | $\nu_\mu$ | 0 |
| $\nu_\tau$ | $8_0$ | $\nu_\tau$ | 0 |
| $\mu$ | $6_0 + 7_0 + 7_1$ | $e + \nu_\mu + \mu_7$ | 105.6 MeV |
| $\tau$ | $6_0 + 7_0 + 7_2$ | $e + \nu_\mu + \tau_7$ | 1786 MeV |
| $\mu'$ | $6_0 + 7_0 + 7_2 + 8_0 + 8_1$ | $e + \nu_\mu + \mu_7 + \nu_\tau + \mu_8$ | 136.9 GeV |
| Quarks | $d_a$ for quarks | | |
| u | $5_0 + 7_0 + 7_1$ | $q_5 + q_7 + u_7$ | 330.8 MeV |
| d | $6_0 + 7_0 + 7_1$ | $q_6 + q_7 + d_7$ | 332.3 MeV |
| s | $6_0 + 7_0 + 7_2$ | $q_6 + q_7 + s_7$ | 558 MeV |
| c | $5_0 + 7_0 + 7_3$ | $q_5 + q_7 + c_7$ | 1701 MeV |
| b | $6_0 + 7_0 + 7_4$ | $q_6 + q_7 + b_7$ | 5318 MeV |
| t | $5_0 + 7_0 + 7_5 + 8_0 + 8_2$ | $q_5 + q_7 + t_7 + q_8 + t_8$ | 176.5 GeV |

The principal dimensional fermion for heavy leptons ($\mu$ and $\tau$) is e and $\nu_e$. Auxiliary dimensional fermion is derived from principal dimensional boson in the same way as Eq. (5) to relate the energies for fermion and boson. For the mass of auxiliary dimensional fermion (AF) from principal dimensional boson (B), the equation is Eq. (13).

$$M_{AF_{d,a}} = \frac{M_{B_{d-1,0}}}{\alpha_a} \sum_{a=0}^{a} a^4 \quad , \tag{13}$$

where $\alpha_a$ = auxiliary dimensional fine structure constant, and a = auxiliary dimension number = 0 or integer. The first term, $\frac{M_{B_{D-1,0}}}{\alpha_a}$, of the mass formula (Eq.(13)) for the auxiliary dimensional fermions is derived from the mass equation, Eq. (5), for the principal dimensional fermions and bosons. The second term, $\sum_{a=0}^{a} a^4$, of the mass formula is for Bohr-Sommerfeld quantization for a charge - dipole interaction in a circular orbit as described by A. Barut [1]. As in Barut lepton mass formula, $1/\alpha_a$ is 3/2. The coefficient, 3/2, is to convert the principal dimensional boson mass to the mass of the auxiliary dimensional fermion in the higher dimension by adding the boson mass to its fermion mass which is one-half of the boson mass. Using Eq. (5), Eq. (13) becomes the formula for the mass of auxiliary dimensional fermions (AF).

$$\begin{aligned} M_{AF_{d,a}} &= \frac{3M_{B_{d-1,0}}}{2} \sum_{a=0}^{a} a^4 \\ &= \frac{3M_{F_{d-1,0}}}{2\alpha_{d-1}} \sum_{a=0}^{a} a^4 \\ &= \frac{3}{2} M_{F_{d,0}} \alpha_d \sum_{a=0}^{a} a^4 \end{aligned} \tag{14}$$

The mass of this auxiliary dimensional fermion is added to the sum of masses from the corresponding principal dimensional fermions (F's) with the same electric charge or the same dimension. The corresponding principal dimensional leptons for u (2/3 charge) and d (-1/3 charge) are $\nu_e$ (0 charge) and e (-1 charge), respectively, by adding –2/3 charge to the charges of u and d [12]. The fermion mass formula for heavy leptons is derived as follows.

$$M_{F_{d,a}} = \sum M_F + M_{AF_{d,a}}$$

$$= \sum M_F + \frac{3M_{B_{d-1,0}}}{2} \sum_{a=0}^{a} a^4 \tag{15a}$$

$$= \sum M_F + \frac{3M_{F_{d-1,0}}}{2\alpha_{d-1}} \sum_{a=0}^{a} a^4 \tag{15b}$$

$$= \sum M_F + \frac{3}{2} M_{F_{d,0}} \alpha_d \sum_{a=0}^{a} a^4 \tag{15c}$$

Eq. (15b) is for the calculations of the masses of leptons. The principal dimensional fermion in the first term is e. Eq. (15b) can be rewritten as Eq. (16).

$$M_a = M_e + \frac{3M_e}{2\alpha} \sum_{a=0}^{a} a^4 , \tag{16}$$

a = 0, 1, and 2 are for e, μ, and τ, respectively. It is identical to the Barut lepton mass formula.

The auxiliary dimensional quarks except a part of t quark are $q_7$'s. Eq.(15c) is used to calculate the masses of quarks. The principal dimensional quarks include $3\nu_\mu$ , 3e, and 3μ., $\alpha_7 = \alpha_w$, and $q_7 = 3\mu$. Eq. (15c) can be rewritten as the quark mass formula.

$$M_q = \sum M_F + \frac{3\alpha_w M_{3\mu}}{2} \sum_{a=0}^{a} a^4 , \tag{17}$$

where a = 1, 2, 3, 4, and 5 for u/d, s, c, b, and a part of t, respectively.

To match $l_8$ ($\nu_\tau$), quarks include $q_8$ as a part of t quark. In the same way that $q_7 = 3\mu$ , $q_8$ involves μ'. μ' is the sum of e, μ, and $\mu_8$ (auxiliary dimensional lepton). Using Eq. (15a), the mass of $\mu_8$ is equal to 3/2 of the mass of $B_7$, which is $Z^0$. Because there are only three families for leptons, μ' is the extra lepton, which is "hidden". μ' can appear only as μ + photon. The pairing of μ + μ from the hidden μ' and regular μ may account for the occurrence of same sign dilepton in the high energy level [11]. The principal dimensional quark $q_8$ = μ' instead of 3μ', because μ' is hidden, and $q_8$ does not need to be 3μ' to be different. Using the equation similar to Eq.(17), the calculation for t quark involves $\alpha_8 = \alpha$ , μ' instead of 3μ for principal fermion, and a = 1 and 2 for $b_8$ and $t_8$, respectively. The hiding of μ' for leptons is balanced by the hiding of $b_8$ for quarks.

The calculated masses are in good agreement with the observed constituent masses of leptons and quarks [12]. The mass of the top quark in Reference [13] is 174.3 ± 5.1 GeV in a good agreement with the calculated value, 176.5 GeV.

## 5. *The composition of hadrons: auxiliary dimensional and hadronic dimensional orbitals*

Auxiliary dimensional orbital is dependent on principal dimensional orbital, so quarks from auxiliary dimensional orbital cannot have independent existences with fractional electric charges and hypercharges. Quarks can exist only as the composites that have integral charges and hypercharges. Thus, another set of dimensional orbitals is required for such composites of quarks. This set of dimensional orbital is hadronic dimensional orbital that also has seven orbitals as in principal dimensional orbital and auxiliary dimensional orbital. Hadronic dimensional orbital is the third set of the three sets of seven orbitals.

Hadronic dimensional orbital is found by P. Polazzi [3]. Polazzi explains the relation between lifetimes and masses in terms of the shell structure in atomic orbital. The noble gases (stable atoms), He, Ne, Ar, Kr, Xe, and Rn have the shell numbers, 1, 2, 3, 4, 5, and 6, respectively. The cube root of the number of electrons for noble gases as a function of the shell numbers is a straight line. He finds that the cube root of the masses for stable hadrons as the function of order of these stable hadrons is a straight line. The stable atoms can be explained by atomic orbital, so the stable hadrons can be explained by the hadronic dimensional orbital.

The ground state of hadronic dimensional orbital starts from d = 6, which is $\pi_{1/2}$, the pseudoscalar quark. For hadronic dimensional orbital, the composite of quarks is 2 $\pi_{1/2}$. The hadronic dimensional orbital number (h) is 0 for 2 $\pi_{1/2}$. The highest orbital (h = 7) is the seventh orbital consisting of the combination of two highest mesonic quarks (b's). For baryon, it is after d = 7, and starts from h = 2. For baryon, the hadron at h = 2 is proton. Fig. 2 shows the hadronic dimensional orbital with the hadronic number, h, and the mesons and the baryons representing the hadronic dimensional orbitals.

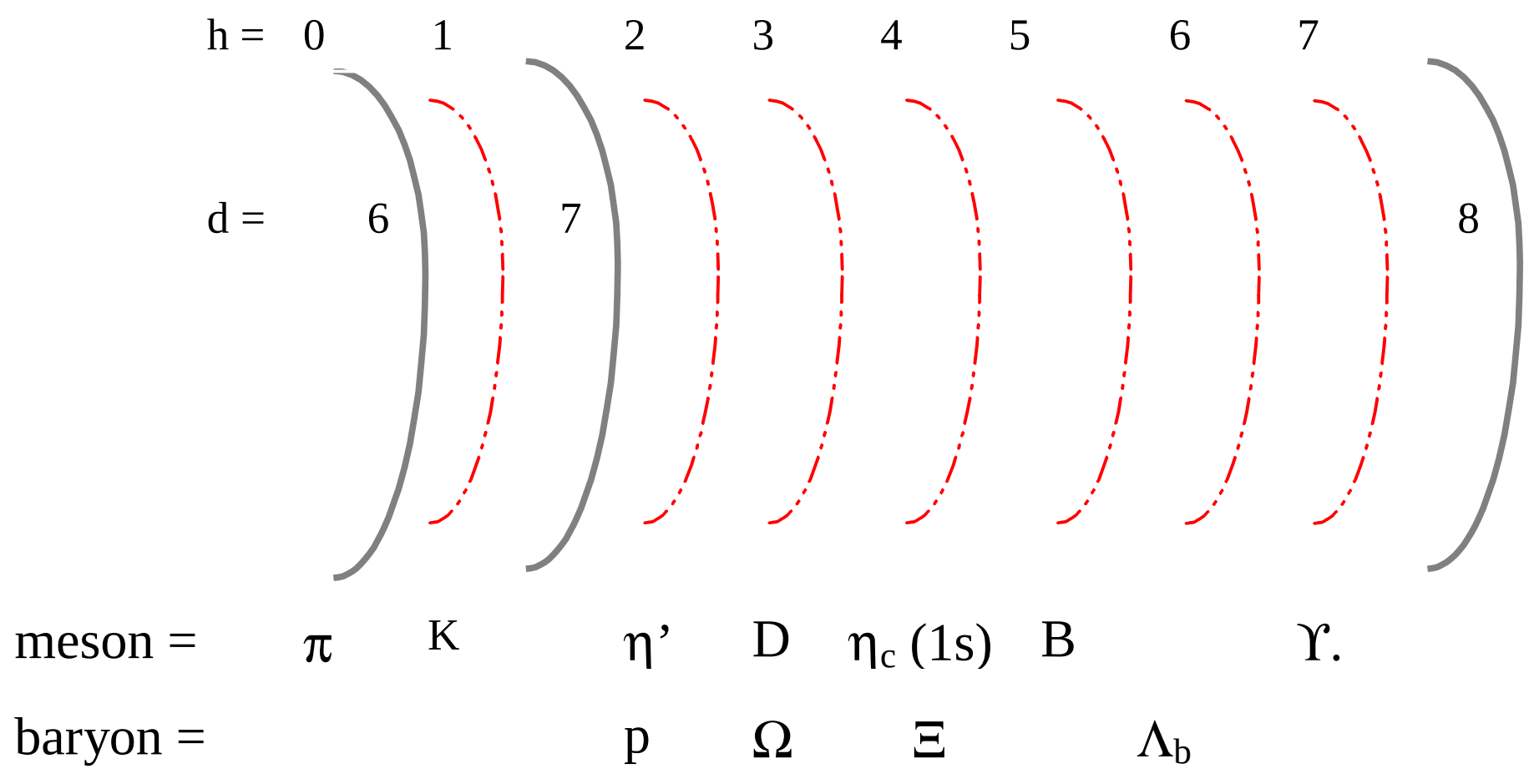

**Fig. 2:** Hadronic dimensional orbital, d = principal dimensional orbital (solid line) number, h = hadronic dimensional orbital (dot line) number

The entity in each orbital is denoted as "hadronic mark". Instead of the fourth power of the auxiliary dimensional orbital number as in the quark mass formula, the mass in the hadronic mark mass formula is the third power of the hadronic number as in atomic orbital.

$$M_h = M_0 \sum_{h=0}^{h} (1 + k\,h)^3, \tag{18}$$

where $M_h$ = the mass of hadronic mark with h hadronic number, $M_0$ is the ground state mass for hadronic mark, h = hadronic number, and k = constant.

Hadronic mark can be mesonic mark (MB) or baryonic mark (BM). For mesons, to connect exactly 140.05 MeV (two $\pi_{1/2}$'s) at h = 0 and 10635.4 MeV (two b quarks) at h = 7, $M_0$ = 140.05 MeV and k = 0.268357. The masses for MM0, MM1, MM2, MM3, MM4, MM5, MM6, and MM7 are 140.05 (two $\pi_{1/2}$'s), 425.8, 934, 1758, 3006, 4805, 7295, and 10635 (two b quarks) MeV, respectively. The stable baryons start at h = 2 with proton, p with the mass of 938.27 MeV near MM2 (934MeV). For baryons, the exact mass for BM4 is 2589.8 MeV as the combined mass for u, s, and c quarks. For baryons, to connect exactly 938.27 MeV (p) for BM2 and 2589.8 MeV ( u, s, and c quarks) for BM4, $M_0$ = 176.098 MeV and k = 0.19095. The masses for BM0, BM1, BM2, BM3, BM4, BM5, BM6, and BM7 are 176.1, 473.6, 938.3 (p), 1624, 2590 (usc), 3905, 5645, and 7891 MeV. The hadronic marks represent the stable hadrons, the significant parts of stable hadrons, and recurrent parts of all hadrons. As shown later. for mesons, MM0, MM1, MM2, MM3, MM4, MM5, and MM7 $\pi$, K, $\eta$', D, J/$\Psi$ and $\eta_c$ (1s), B, and $\Upsilon$, respectively. For baryons, BM2, BM3, BM4, and BM6 represent p, $\Omega$, $\Xi$, and $\Lambda_b$, respectively. The 1/3 power of mass as the function of the hadronic number is a straight line as shown in Fig. 3.

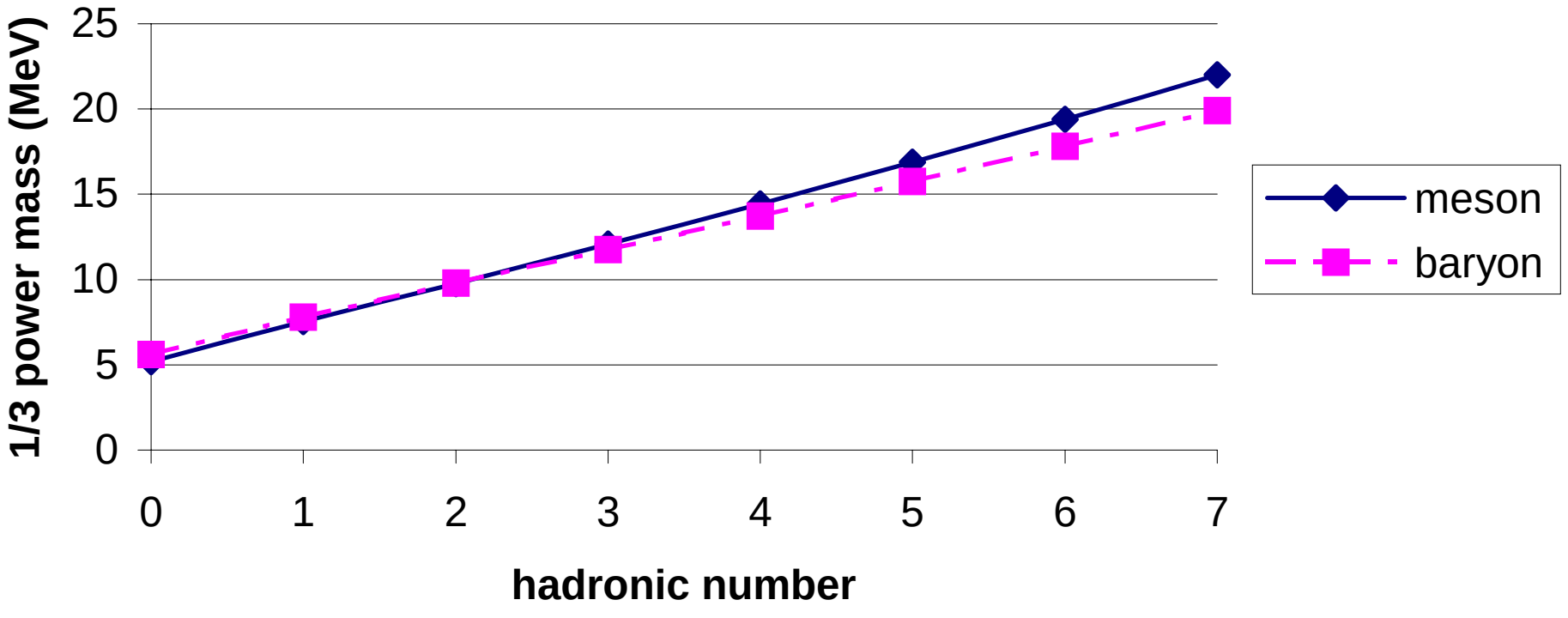

**Fig. 3:** Hadronic dimensional numbers and 1/3 power of hadron masses

Quarks from auxiliary dimensional orbital and hadronic marks from hadronic dimensional orbital constitute the quark-mark formula for hadrons. The quark-mark formula provides the criteria for stable hadrons.

In atomic orbitals, the transition from one orbital to another orbital is through the emission or absorption of gauge boson, photon. The transitions among hadrons are also through gauge bosons, such as weak bosons, gluon, and photon. In addition to these gauge bosons, the transitions among hadrons require the addition and the subtraction of "mass blocks". For example, the transition from proton, p (uud) to $\Lambda^0$ (uds), is through the addition of mass blocks to p. Conversely, the transition from proton, $\Lambda^0$ (uds) to p (uud), is through the subtraction of mass blocks from $\Lambda^0$.

$$p \xleftrightarrow{mass\,block} \Lambda^0$$
$$\Lambda^0 = p + mass\,block \quad (17)$$

The composition of $\Lambda^0$ is p + mass block. Such formula involving mass blocks is the mass block formula.

All hadrons, including proton and neutron, have the mass block formulas, so all hadrons share the common mass blocks. In terms of mass, the quark-mark represents hadron only approximately. The mass from the quark-mark formula is matched by the mass from the mass block formula, which represents the mass of hadron precisely.

$$M_{quark-mark\ \ formula} \approx M_{mass\,block\ \ formula} = M_{hadron} \quad (20)$$

All hadrons are represented by the mass block formula. Mass blocks consist of the lowest- mass quark ($\pi_{1/2}$ and u), lepton (e), and baryons (p and n). The mass block formula is parallel to MacGregor-Akers constituent quark model [4][5], whose calculated masses and the predicted properties of hadrons are in very good agreement with observations. In the constituent quark model, the mass building blocks are the "spinor" (S with mass 330.4 MeV) and the mass quantum (mass = 70MeV). For the mass block formula, S, corresponds to u quark with the lowest mass (330.77 MeV), which is designated as U mass block. The basic quantum is the pseudoscalar quark, $\pi_{1/2}$ (mass = 70.025 MeV), which is denoted as m mass block. The mass of MM1 is 425MeV, which is close to six m's (420MeV), so six m's appear together as a group that is denoted as X, which is not a fundamental mass block.

In additional to U and m, the mass block formula includes P (positive charge) and N (neutral charge) with the masses of proton and neutron. As in the constituent quark model, the mass associated with positive or negative charge is the electromagnetic mass, 4.599 MeV, which is nine times the mass of electron. This mass (nine times the mass of electron) is derived from the baryon-like electron that represents three quarks in a baryon and three electrons in $d_6$ quark as in Table 2. This electromagnetic mass is the baryonic electromagnetic mass. This electromagnetic mass is observed in the mass difference between $\pi^\circ$ ($m_2$) and $\pi^+$ ($m_2e$) where e denotes electromagnetic charge. The calculated mass different is one electromagnetic mass, 4.599 MeV, in good agreement with the observed mass difference, 4.594 MeV, between $\pi^\circ$ and $\pi^+$. (The values for observed masses are taken from "Particle Physics Summary "[13].) The particles in the mass block formula are listed in Table 4.

**Table 4.** Mass blocks in the mass block formula

| Blocks | m | X* | U | N | P | Electromagnetic mass |
|---|---|---|---|---|---|---|
| Origins in quarks | $\pi_{1/2}$ | 6 $\pi_{1/2}$ | u quark | n | p | 9 e |
| Mass (MeV) | 70.0254 | 420.2 | 330.77 | 939.565 | 938.272 | 4.599 |

*X is MM1, which is not a fundamental mass block

The mass block formula can be represented by two different formulas: the initial mass block formula and the decay mass block formula. The initial formula is the initial match to the quark-mark formula. The initial formula reflects the structure of quark-mark formula. The decay formula is the match to the decay mode. The dominated fraction of the decay modes typically reflects the most stable hadron and the most favorable decay path. The decay mass block formula consists of the largest hadron in the dominated fraction and mass blocks. In some cases, the initial formulas are identical to the decay formula. For example, in terms of quark-mark formula, $\Lambda^0$ (uds) is p (uud) + BM0 (baryonic mark 0), representing the transition of u quark to s quark. This quark-mark formula is matched by the initial formula, $pm_3e$, which means p + three m's + e. The initial formula is identical to the decay formula that represents p in the dominated fraction of decay mode. In some cases, the initial formula is not same as the decay formula. For example, in terms of the quark-mark formula, $\Sigma^+$(1385) (uus) is $\Sigma^+$ + BM0, representing the transition from $J^p = ½^+$ to $J^p = 3/2^+$. This quark-mark formula is matched by the initial formula, $\Sigma^+m_3$. The decay formula is $\Lambda^\circ m_4e$, where the dominated decay mode involves $\Lambda^\circ$. The calculated masses by the initial formula and the decay formula are 1384.8 MeV and 1385.7 MeV, respectively, comparing the observed mass of 1382.8 MeV. In few cases, the only mass block formula is the initial mass block formula. For example, the mass block formula for n is $U_3$. There is no decay formula that can calculate the mass of n. Such cases involve U.

Hadrons are the composites of mass blocks as molecules composing of atoms. As atoms are bounded together by chemical bonds, mass blocks are bounded by "hadronic bonds," connecting the mass blocks in the mass block formula. These hadronic bonds are similar to the hadronic bonds in the constituent quark model.

The hadronic bonds are the overlappings of the auxiliary dimensional orbitals. From Eq (13), the energy for the auxiliary orbital for U (u quark) is

$$\begin{aligned} E_a &= (3/2)\,(3\,M_\mu)\,\alpha_w \\ &= 14.122 \text{ MeV} \end{aligned} \tag{21}$$

The auxiliary orbital is a charge - dipole interaction in a circular orbit as described by A. O. Barut [1], so a fermion for the circular orbit and an electron for the charge are embedded in this hadronic bond. The fermion for the orbital is the supersymmetry fermion for the auxiliary dimensional orbital according to Eq (5).

$$M_f = E_a\,\alpha_w \tag{22}$$

The binding energy (negative energy) for the bond (U-U) between two U's is twice of 14.122 MeV minus the masses of the supersymmetry fermion and electron.

$$\begin{aligned} E_{U\text{-}U} &= -\,2\,(E_a - M_f - M_e) \\ &= -26.384 \text{ MeV} \end{aligned} \tag{23}$$

It is similar to the binding energy (-26 MeV) in the constituent quark model. An example of U─U bond is in neutron (U – U – U) which has two U – U bonds. The mass of neutron can be calculated as follows.

$$\begin{aligned} M_n &= 3M_U + 2E_{U\text{-}U} \\ &= 939.54 \text{ MeV}, \end{aligned} \tag{24}$$

which is in excellent agreement with the observed mass, 939.57 MeV. The mass of proton is the mass of neutron minus the mass difference (three times of electron mass = $M_{3e}$) between u and d quark as shown in Table 2. Proton is represented as U – U – (U - 3e). The calculation of the mass of proton is as follows.

$$\begin{aligned} E_a \text{ for (U-3e)} &= (3/2)\,(3\,(M_\mu - M_{3e}))\,\alpha_w \\ M_f &= E_a\,\alpha_w \\ M_p &= 2\,M_U + M_{(U\text{-}3e)} + E_{U\text{-}U} + E_{U\text{-}(U\text{-}3e)} \\ &= 938.21 \text{ MeV} \end{aligned} \tag{25}$$

The calculated mass is in a good agreement with the observed mass, 938.27 MeV.

The binding energy for m – m ($\pi_{1/2} - \pi_{1/2}$) bond can be derived in the same way as Eqs (21), (22), and (23).

$$E_a = (3/2)\, M_m\, \alpha_w$$
$$M_f = E_a\, \alpha_w$$
$$E_{m-m} = -2\,(E_a - M_f - M_e) = -5.0387 \text{ MeV} \qquad (26)$$

It is similar to the binding energy (-5 MeV) in the constituent quark model. An example for the binding energy of m - m bond is in $\pi^\circ$. The mass of $\pi^\circ$ can be calculated as follows.

$$M_{\pi^\circ} = 2\, M_m + E_{m-m} = 135.01 \text{ MeV}. \qquad (27)$$

The calculated mass of $\pi^\circ$ is in excellent agreement with the observed value, 134.98 MeV. There is one m - m bond per pair of m 's, so there are two m - m bonds for 4 m 's, and three m - m bonds for 6 m 's.

Another bond is N – m or P – m, the bond between neutron or proton and m. Since N is UUU, N – m bond is derived from U – m. The binding energy of U – m is the average between U–U and m - m.

$$E_{U-m} = 1/2\,(E_{U-U} + E_{m-m}) \qquad (28)$$

An additional dipole ($e^+e^-$) is needed to connected U – m to neutron.

$$E_{N-m} = E_{U-m} + 2\, M_e = -14.689 \text{ MeV}. \qquad (29)$$

It is similar to -15 MeV in the constituent quark model. An example for N–m is $\Sigma^+$ which is represented by $pm_4$ whose structure is $m_2$ –p–$m_2$. The 4 m 's are connected to p with two N–m bonds. The mass of $\Sigma$ is as follows.

$$M_{\Sigma+} = M_n + 4\, M_m + 2\, E_{N-m} = 1189.0 \text{ MeV}.$$

The observed mass is 1189.4 MeV.

N-m bond can be generalized for the transformations of all baryons. The numbers of N-m bond is quantized by $J^p$ and the type of quarks during the baryonic transformation as follows.

**Table 5**: Quantized N-m bond

| | number of N-m bonds |
|---|---|
| <u>change $J^P$</u> | |
| $1/2^+ \rightarrow 3/2^+$ | + 1 |
| $1/2^+ \rightarrow 1/2^-$ | 0 |
| $1/2^+ \rightarrow 3/2^-$ | - 1 |
| <u>change quark</u> | |
| $u / d \rightarrow s$ at $J^P = 1/2^+$ | + 2 |
| $s \rightarrow c$ | + 1 |
| $c \rightarrow b$ | + 1 |

There is N – N hadronic bond between two N’s. N has the structure of U – U – U. N – N has a hexagonal structure shown in Fig. 4.

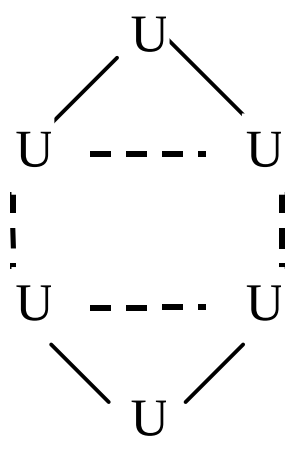

**Fig. 4:** The structure of N - N

There are two additional U – U for each N. The total number of U – U bonds between two N’s is 4. An example is J/ψ ($c\bar{c}$) has the mass block formula of $N_2U_4$ with the structure. The mass of J/ψ can be calculated as follows.

$$
\begin{aligned}
M_{J/\psi} &= 2M_N + 4M_U + 4\,E_{U-U} \\
&= 3096.7 \text{ MeV}
\end{aligned}
$$

The calculated mass for J/ψ is in a good agreement with the observed mass (3096.9 MeV).

Among the particles in the mass block formula, there are hadronic bonds, but not all particles have hadronic bonds. In the mass block formula, hadronic bonds appear only among the particles that relate to the particles in the corresponding quark-mark formula. The unrelated particles have no hadronic bonds. In the mass block formula, for baryons other than p and n, the core particles are P, N, and m. For the mesons consisting of u and d quarks, the core particles are m, U, and N. For the mesons containing one u, d, or s along with s, c, or b, the core particles are U and N, and no hadronic bond exist among m's. For the mesons ($c\bar{c}$ and $b\bar{b}$), the only hadronic bond is N – N. The occurrences of hadronic bonds are listed in Table 6.

**Table 6.** Hadronic bonds in hadrons

| | U –U | m – m | N(P) – m | N –N |
|---|---|---|---|---|
| Binding energy (MeV) | -26.384 | -5.0387 | -14.6894 | 2 U–U per N |
| Baryons other than n and p | | | √ | √ |
| Mesons with u and d only | √ | √ | | √ |
| Mesons containing one u, d, or s along with s, c, or b | √ | | | √ |
| c $\bar{c}$ or b $\bar{b}$ mesons | | | | √ |

An example is the difference between π and $f_0$. The decay modes of $f_0$ include the mesons of s quarks from K meson. Consequently, there is no m-m for $f_0$. The mass block formula for $f_0$ is $\pi^0 X_2$. Since $\pi^0$ has one m-m bond, the calculation involves the subtraction of this m-m bond. The mass of $f_0$ is as follows.

$$\begin{aligned} M &= M\pi^0 + 12\, M_m - E_{m-m} \text{ from } \pi^0 \\ &= 980.3 \text{ MeV} \end{aligned}$$

The observed mass is 980 ± 10 MeV.

In additional to the binding energies for hadronic bonds, hadrons have Coulomb energy (-1.2 MeV) between positive and negative charges and magnetic binding energy (±2MeV per interaction) for U–U from the constituent quark model [4]. In the constituent quark model, the dipole moment of a hadron can be calculated from the magnetic binding energy. Since in the mass block formula, magnetic binding energy becomes a part of hadronic binding energy as shown in Eq (23), magnetic binding energy for other baryons is the difference in magnetic binding energy between a baryon and n or p. If a baryon has a similar dipole moment as p or n, there is no magnetic binding energy for the baryon. An example for Coulomb energy and magnetic binding energy is Λ (uds, J=1/2) whose formula is $Pm_3$- with the structure of $m_2$–P–$m_1$- e. The dipole moment of $\Lambda^0$ is –6.13 $\mu_N$, while the dipole moment of proton (P) is 2.79 $\mu_N$ [4]. According to the constituent quark model, this difference in dipole moment represents -6 MeV magnetic binding energy. The Coulomb energy between the positive charge P and the negative charge is –1.2 MeV. The electromagnetic mass for is 4.599 MeV. The mass of $\Lambda^0$ is calculated as follows.

$$\begin{aligned} M_{\Lambda}{}^0 &= M_P + 3Mm + M_{e.m.} + 2\, E_{N-m} + E_{mag} + E_{coul} \\ &= 1116.4 \text{ MeV} \end{aligned}$$

The observed mass is 1115.7 ± 0.0006 MeV.

The quark-mark formula involves both quarks and marks. The formula constitutes the approximated mass for hadrons. Table shows the quark-mark formula for some baryons.

**Table 7** : The Quark-Mark Formulas for Baryons

| Quark-Mark | MeV (calculated) | Baryon |
|---|---|---|
| Mark | | |
| *BM0* | 176.1 | d/u $\rightarrow$ s |
| | | p (udd) + BM0 ≈ $\Lambda^\circ$ (uds) |
| | | p (udd) + 2BM0 ≈ $\Xi^o$ (uss) |
| | | $\Lambda^+_c$ (udc) + BM0 ≈ $\Xi^+_c$ (usc) |
| | | $\Xi^+_c$ (usc) + BM0 ≈ $\Omega_c^0$ (ssc) |
| | | $\rightarrow$ higher $J^p$ with same quarks |
| | | $\Lambda^+_c$ ($1/2^+$) + BM0 ≈ $\Sigma_c^+$ ($1/2^+$) |
| *BM1* | 473.6 | $\rightarrow$ higher $J^p$ with same quarks |
| | | $\Xi^\circ$ ($1/2^+$) + BM1 ≈ $\Xi^\circ$ (1820)($3/2^-$) |
| *BM2* | 938.27 | ≈ p (uud) |
| *BM3* | 1623.5 | ≈ $\Omega^-$ (sss) (also ≈ s + s + s) |
| *BM4* | 2589.8 | ≈ $\Xi'^+_c$ (usc) (also u + s + c) |
| *BM5* | 3905.1 | not found |
| *BM6* | 5644.7 | ≈ $\Lambda^\circ_b$ (udb) |
| Quark | 70.02 | m = $\pi_{1/2}$ |
| | | $\rightarrow$ higher $J^p$ with same quarks |
| | | $\Sigma_c$ $(2455)^+$ ($1/2^+$) + m $\rightarrow$ $\Sigma_c$ $(2520)^+$($3/2^+$) |
| | 1219.5 | u + u + s ≈ $\Sigma^+$ (uus) |
| | 1222.6 | d + d + s ≈ $\Sigma^-$ (dds) |
| | 2309.8 | n (udd) + c – u ≈ $\Lambda^+_c$ (udc) |
| | 2685.0 | $\Xi^o$(uss) + c – u ≈ $\Omega^o$ (ssc) |

BM0 represents the transition from u/d $\rightarrow$ s and transition to different $J^p$ with the same quarks. BM1 represents the transition to different $J^p$ with the same quarks. Another way for the transition to different $J^p$ with the same quark is through m. BM2, BM3, BM4, and BM6 corresponding to p, $\Omega^-$, $\Xi'^+_c$, and $\Lambda^\circ_b$, respectively. The masses of $\Sigma^+$ (uus), $\Sigma^-$ (dds), $\Lambda^+_c$ (udc), $\Omega^o_c$ (ssc) are about the masses of u + u + s, d +d + s, n(udd) + c – u, and $\Xi^o$(uss) + c – u, respectively.

Such quark-mark formulas can be matched with the initial mass block formulas. The initial mass block formulas in the same group can be connected together into a the initial mass block sequence. For example, for the $\Lambda/\Xi$ group, the initial mass block sequence is as Fig. 5.

$$\Sigma^{+}(1385)$$

$$\uparrow BB0 \approx m_3$$

$$p \xrightarrow{uus - uud \,\approx\, m_4} \Sigma^{+} \xrightarrow{e} \Sigma^{0} \xrightarrow{BB0 \,\approx\, m_3} \Sigma^{0}(1385) \xrightarrow{BB0 \,\approx\, m_3} \Lambda^{0}(1405)$$

$$\downarrow BB0 \approx m_3 e$$

$$\Lambda^{0}$$

$$n \xrightarrow{dds - uud \,\approx\, m_4 e} \Sigma^{-} \xrightarrow{BB0 \,\approx\, m_3} \Sigma^{-}(1385)$$

$$\downarrow BB0 + BB1 \approx m_8 e_2$$

$$\Lambda^{0}(1520)$$

**Fig. 5:** The initial mass block sequence for Λ/Σ baryons

The decay mass block formula involves the mass blocks plus the largest baryon in the dominated fraction of decay modes. The calculated masses from the initial formula and the decay formula for the same hadron are nearly the same. Table and Table show $I(J^P)$, quark compositions, the initial mass block formulas derived from the quark-mark formulas, the decay mass block formulas, the calculated masses based on the decay mass block formulas, and the comparisons with the observed masses for all well-defined baryons.

**Table 8.** Baryons (N, $\Lambda/\Sigma$, Ξ°, and $\Lambda_c/\Sigma_c$ baryons)

| Baryons | I($J^P$ ) | quark | initial formula and sequence | decay formula | Calculated mass | Observed mass | Difference |
|---|---|---|---|---|---|---|---|
| N-baryons | | | | | | | |
| n | ½(½$^+$) | udd | $U_3$ | $U_3$ | 939.54 | 939.57 | -0.03 |
| p | ½(½$^+$) | uud+ | BM2 = $U_3$-1/3e | $U_3$-1/3e | 938.21 | 938.27 | -0.06 |
| $\Lambda/\Sigma$ baryons | | | | | | | |
| Λ° | 0(½$^+$) | uds | p + BM0 ≈ $pm_3e$ | $pm_3e$ | 1116.4 | 1115.7 | 0.7 |
| $\Sigma^+$ | 1(½$^+$) | uus+ | u + u + s ≈ $pm_4$ | $pm_4$ | 1189.0 | 1189.4 | -0.4 |
| Σ° | 1(½$^+$) | uds | u + d + s ≈ $\Sigma^+e$ | $\Lambda°me_2$ | 1193.7 | 1192.6 | 1.1 |
| $\Sigma^-$ | 1(½$^+$) | dds- | d + d + s ≈ $nm_4e$ | $nm_4e$ | 1194.9 | 1197.4 | -2.5 |
| $\Sigma^+$(1385) | 1(3/2$^+$) | uus+ | $\Sigma^+$ + BM0 ≈ $\Sigma^+m_3$ | $\Lambda°m_4e$ | 1385.7 | 1382.8 | 2.9 |
| Σ°(1385) | 1(3/2$^+$) | uds | Σ° + BM0 ≈ $\Sigma°m_3$ | $\Lambda°m_4$ | 1381.1 | 1383.7 | -2.6 |
| $\Sigma^-$(1385) | 1(3/2$^+$) | dds- | $\Sigma^-$ + BM0 ≈ $\Sigma^-m_3$ | $\Lambda°m_4e$ | 1385.7 | 1387.2 | -1.5 |
| Λ°(1405) | 0(½-) | uds | Σ°+ BM0 ≈ $\Sigma°m_3$ | $\Sigma°m_3$ | 1402.7 | 1406.0 | 3.3 |
| Λ° (1520) | 0(3/2$^-$) | uds | n + BM0 + BM1≈ $nm_8e_2$ | $nm_8e_2$ | 1522.5 | 1519.5 | 3.0 |
| Ξ° baryons | | | | | | | |
| Ξ° | ½(½$^+$)[a] | uss | p + 2BM0 ≈ $pm_6e$ | $\Lambda°m_3e_2$ | 1313.1 | 1314.8 | -1.7 |
| $\Xi^-$ | ½(½$^+$)[a] | dss | Ξ° e | $\Lambda°m_3e_3$ | 1322.5 | 1321.3 | 1.2 |
| Ξ°(1530) | ½(3/2$^+$) | uss | Ξ° + BM0 ≈ $\Xi°m_3e_2$ | $\Xi°m_3e_2$ | 1532.9 | 1531.8 | 1.1 |
| $\Xi^-$(1530) | ½(3/2$^+$) | dss- | $\Xi^0$ (1530)e | $\Xi°m_3e_3$ | 1536.3 | 1535.0 | 1.3 |
| Ξ°(1820) | ½(3/2$^-$) | uss | Ξ° + BM1 ≈ $\Xi°m_7$ | $\Lambda°m_{10}$ | 1830.6 | 1823.0 | 7.6 |
| $\Lambda_c/\Sigma_c$ baryons | | | | | | | |
| $\Lambda^+_c$ | 0(½$^+$)[a] | udc+ | p + c – u ≈ $p_2m_8e_3$ | $p_2m_8e_3$ | 2283.9 | 2284.9 | -1.0 |
| $\Sigma_c^+$(2455) | 1(½$^+$) | udc+ | $\Lambda^+_c$ +BM0≈ $\Lambda^+_cm_2e_8$ | $\Lambda^+_cm_2e_8$ | 2453.3 | 2451.3 | 2.0 |
| $\Sigma_c^0$(2455) | 1(½$^+$) | ddc | $\Sigma_c^+$(2455)e | $\Lambda^+_cm_2e_9$ | 2456.7 | 2452.2 | 4.5 |
| $\Sigma_c^{++}$(2455) | 1(½$^+$)[a] | uuc++ | $\Sigma_c^+$(2455)e | $\Lambda^+_cm_2e_8$ | 2456.7 | 2452.6 | 4.1 |
| $\Sigma_c^+$(2520) | 1(3/2$^+$)[a] | udc+ | $\Sigma_c^+$(2455)m | $\Lambda^+_cm_3e_{10}$ | 2515.5 | 2515.9 | -0.4 |
| $\Sigma_c^0$(2520) | 1(3/2$^+$)[a] | ddc | $\Sigma_c^0$(2455)m | $\Lambda^+_cm_3e_{11}$ | 2518.9 | 2517.5 | 1.4 |
| $\Sigma_c^{++}$(2520) | 1(3/2$^+$)[a] | uuc++ | $\Sigma_c^{++}$(2455)m | $\Lambda^+_cm_3e_{11}$ | 2518.9 | 2519.4 | -0.5 |
| $\Lambda^+_c$(2593) | 0(½$^-$) | udc+ | $\Sigma_c^0$(2520)m | $\Lambda^+_cm_4e_8$ | 2593.4 | 2593.9 | -0.5 |
| $\Lambda^+_c$(2625) | 0(3/2$^-$)[a] | udc+ | $\Lambda^+_c$ +2BM0 ≈$\Lambda^+_cm_2e_8$ | $\Lambda^+_cm_2e_8$ | 2628.5 | 2626.6 | 1.9 |

a = $J^p$ that is predicted by quark model, not measured

**Table 9.** Baryons ($\Omega$, $\Xi_c$, $\Omega_c$, and $\Lambda_b$ baryons)

| Baryons | $I(J^P)$ | Quark | initial formula and sequence | decay formula | Calculated mass | Observed mass | Difference |
|---|---|---|---|---|---|---|---|
| $\Omega$ baryons | | | | | | | |
| $\Omega^-$ | $0(3/2^+)^a(3/2^-)^b$ | sss | BM3 $\approx$s + s + s | $\Lambda° m_8 e_3$ | 1672.6 | 1672.5 | 0.1 |
| $\Xi_c$ baryons | | | | | | | |
| $\Xi^+_c$ | $½(½^+)^a(3/2^+)^b$ | usc+ | $\Lambda^+_c$ +BM0 $\approx \Lambda^+_c m_3$ | $\Xi° m_{17} e$ | 2465.8 | 2466.3 | -0.5 |
| $\Xi°_c$ | $½(½^+)^a(3/2^+)^b$ | dsc | $\Xi^+_c$ e | $\Xi^+ m_{17} e$ | 2472.3 | 2471.8 | 0.5 |
| $\Xi'^+_c$ | $½(½^+)^a$ | usc+ | BM4 =u+s +c $\approx \Xi^+_c me_6$ | $\Xi^+_c me_6$ | 2572.6 | 2574.1 | -1.5 |
| $\Xi'°_c$ | $½(½^+)^a$ | dsc | $\Xi'^+_c e$ | $\Xi^0_c me_{10}$ | 2578.1 | 2578.8 | -0.7 |
| $\Xi^+_c$ (2645) | $½(3/2^+)^a(1/2^+)^b$ | usc+ | $\Xi^+_c$+BM0 $\approx \Xi^+_c m_2 e_8$ | $\Xi^+_c m_2 e_8$ | 2649.4 | 2647.4 | 2.0 |
| $\Xi°_c$(2645) | $½(3/2^+)^a(1/2^+)^b$ | dsc | $\Xi°_c$+BM0 $\approx \Xi°_c m_2 e_8$ | $\Xi^+_c m_2 e_7$ | 2646.0 | 2644.5 | 1.5 |
| $\Xi^+_c$(2790) | $½(½^-)^a$ | usc+ | $\Xi'^+_c$+BM0 $\approx \Xi'^+_c m_3 e_2$ | $\Xi'^+_c m_3 e_2$ | 2792.2 | 2790.0 | 2.2 |
| $\Xi°_c$(2790) | $½(½^-)^a$ | dsc | $\Xi'^+_c$ + BM0 $\approx \Xi'°_c m_3$ | $\Xi'°_c m_3$ | 2788.9 | 2790.0 | -1.1 |
| $\Xi^+_c$ (2815) | $½(3/2^-)^a(3/2^+)^b$ | usc+ | $\Xi^+_c$ + 2BM0 $\approx \Xi^+_c m_5$ | $\Xi^+_c m_5$ | 2816.4 | 2814.9 | 1.5 |
| $\Xi°_c$(2815) | $½(3/2^-)^a(3/2^+)^b$ | dsc | $\Xi°_c$ + 2BM0 $\approx \Xi^0_c m_5$ | $\Xi^0_c m_5$ | 2821.9 | 2819.0 | 2.9 |
| $\Omega_c$ baryons | | | | | | | |
| $\Omega°_c$ | $0(½^+)^a$ | ssc | $\Xi^+_c$ + BM0 $\approx$ $\Xi°$ + c – u | $\Omega^- m_{15} e$ | 2696.9 | 2697.5 | -1.1 |
| $\Lambda_b$ baryons | | | | | | | |
| $\Lambda°_b$ | $0(½^+)^a(3/2^+)^b$ | udb | BM6 $\approx$ b + 2BM0 | $\Lambda^+_c m_{48} e$ | 5621.3 | 5624.0 | -2.7 |

a = $J^p$ that is predicted by quark model, not measured
b = $J^p$ that fits the calculation

For mesons, mesonic marks are matched and replaced by mass blocks, such as X and U. MM1, MM2, MM3, MM4, MM5, MM6, and MM7 are replaced by X, $U_3$, $U_6$, $U_9$, $U_{16}$, $U_{24}$, and $U_{32}$ with masses 420.1, 939.5, 1853, 2977, 4897, 7332, and 10587 MeV respectively, which are close to the original masses, 425.8, 934, 1758, 3006, 4805, 7295, 10635 MeV, respectively.

$\pi$ represents MM0. MM1 is X, and Xme = $K^\pm$. X is also for the transition between two different hadrons with the same quarks. U and m are also for the transition between two different hadrons with the same quarks. $\eta$'(958) represents MM2. MM2 is also $U_3$, MM2that is one of the mass blocks in the mass block formula. K represents MM1. MM2 is $U_3$ and $\phi$ (1020) = $U_3 me_2$, analogous to Xme for $K^\pm$. MM3 is $U_6$, and $U_6 e_2$ = D(1865)°. Both $\eta_c$ (1s) and J/$\psi$ represent MM4 for c $\bar{c}$ mesons. Possibly, due to the absence of MM6 ($U_{24}$), MM5 ($U_{16}$) moves toward higher mass, resulting in $U_{17} me_3 = B^0$ that is higher mass than $U_{16}$. The absence of MM6 also allows a number of stable $\Upsilon$'s (b $\bar{b}$) below MM7. MM7 ($U_{32}$) is $\Upsilon$ (4s) that is the only $\Upsilon$ whose decay mode is $B^0$ $\bar{B}^0$ that is MM5. Table 10 shows the quark-mark formulas for mesons.

**Table 10:** The Quark-mark Formula for Mesons

| Quark-Mark | mass block | MeV (calculated) | Mesons |
|---|---|---|---|
| Mark | | | |
| *MM0* | $m_2$ | 140.05 | u/d with m-m bond ($\pi = m_2$) |
| *MM1* | X | 420.15 | Xm Xme = $K^{\pm}$ |
| | | | $\rightarrow$ different hadron with the same quarks |
| | | | $\pi^0$ ($1(0^{-+})$) + X $\rightarrow \eta$ ($0(0^{-+})$) |
| *MM2* | $U_3$ | 939.5 | $\approx \eta$ '(958) |
| | | | $U_3me_2 = \phi$ (1020) |
| *MM3* | $U_6$ | 1852.7 | $U_6e_2 = \overline{D}(1865)^{\circ}$ |
| *MM4* | $U_9$ | 2976.9 | $U_9 = c\ \bar{c}\eta_c$ (1s) |
| | $U_4N_2$ | 3096.7 | $U_4N_2 = J/\psi$ |
| *MM5* | $U_{16}$ | 4896.6 | $U_{17}me_2 = B^{\pm}$ |
| *MM6* | $U_{24}$ | 7332 | none |
| *MM7* | $U_{32}$ | 10584 | $\Upsilon$ (4s) (also b + b) |
| Quark | | $m_n$ | $\rightarrow$ different hadron with same quarks |
| | | 70.03 (m) | $\eta$ (1405) + m $\rightarrow \eta$ (1475) |
| | | $U_n$ | $\rightarrow$ different hadron with the same quarks |
| | | 330.67 (U) | $\eta$ '(958) ($0(0^{-+})$) + U $\rightarrow \eta$ (1295) (($0(0^{-+})$) |
| | | 635.2 ($U_2$) | $\pi^0$ ($1(0^{-+})$) + $U_2 \rightarrow \rho$ (770) ($1(0^{--})$) |
| | | 1008.3 | p + m $\approx K^*(892)^{\pm}$ |
| | | 888.8 | u + s $\approx \phi$ (1020) |
| | | 890.3 | d + s $\approx K^*(892)^{o}$ |
| | | 2031.8 | u + c $\approx D^*(2007)^{o}$ |
| | | 2033.3 | d + c $\approx D^*(2010)^{\pm}$ |

The quark-mark formulas can be matched with the initial mass block formulas. The initial mass block formulas in the same group can be connected together into the initial mass block sequence. For example, for the unflavored group with only u/d, the initial mass block sequence is as Fig. 6.

$$\pi(140) \qquad\qquad \pi_1(1400)$$

$$\uparrow e \qquad\qquad \uparrow X_2 e_2$$

$$\pi(135) \xrightarrow{Xe_2} \eta(547) \xrightarrow{X} \eta'(958) \xrightarrow{U} \eta(1295) \xrightarrow{e} \pi(1300)$$

$$\downarrow U_2$$

$$\rho(770) \xrightarrow{X} f_1(1170) \xrightarrow{m_2U} \omega(1650)$$

$$\downarrow e_2$$

$$\omega(780) \xrightarrow{U_2} \omega(1420)$$

**Fig. 6:**. The initial mass block sequence for the unflavored group with only u/d

Tables 11, 12, and 13 show I($J^P$), quark compositions, the initial mass block formulas derived from the quark-mark formulas, the decay mass block formulas, the calculated masses based on the decay mass block formulas, and the comparisons with the observed masses for all well-defined mesons.

## **Table 11.** Unflavored mesons

| Meson | $I^G (J^{pc})$ | Quark | initial formula and sequence | Decay formula | Calculated mass. | Observed mass | Difference |
|---|---|---|---|---|---|---|---|
| <u>only u, d, or lepton in decay mode</u> | | | | | | | |
| $\pi°$ (135) | $1^- (0^{-+})$ | u/d | MM0 | $m_2$ | 135.0 | 134.98 | 0.04 |
| $\pi^\pm$ (140) | $1^- (0^-)$ | u/d | MM0 | $m_2e$ | 139.6 | 139.57 | 0.04 |
| $\eta$ (547) | $0^+ (0^{-+})$ | u/d | $\pi$ + MM1≈$\pi°Xe_2$ | $\pi°Xe_2$ | 548.0 | 547.3 | 0.7 |
| $\eta$'(958) | $0^+ (0^{-+})$ | u/d | $\pi$ + 2MM1 ≈ MM2 | $\eta$ (547)$Xe_2$ | 960.3 | 957.8 | 2.5 |
| $\eta$ (1295) | $0^+ (0^{-+})$ | u/d | $\eta$'(958) U | $\eta$ (547)$UXe_2$ | 1291.1 | 1293.0 | -1.9 |
| $\pi$ (1300) | $1^- (0^{-+})$ | u/d | $\eta$'(958)$Ue_2$ | $\rho$ (770)$Um_3$ | 1301.9 | 1300.0 | 1.9 |
| $\rho$ (770) | $1^+ (1^{--})$ | u/d | $\pi$ + d + u ≈ $\pi^0$ $U_2$ | $\pi°$ (135)$U_2$ | 770.1 | 771.1 | -1.0 |
| $\omega$ (782) | $0^- (1^{--})$ | u/d | $\rho$ (770)$e_2$ | $\pi°$ (135)$U_2e_2$ | 778.1 | 782.6 | -4.5 |
| $h_1$ (1170) | $0^- (1^{+-})$ | u/d | $\rho$ (770) X | $\rho$ (770) X | 1176.1 | 1170.0 | 6.1 |
| $\pi_1$ (1400) | $1^-(1^{-+})$ | u/d | $\eta$ (547) $X_2$ $e_2$ | $\eta$ (547)$X_2$ $e_2$ | 1365.4 | 1370.0 | -4.6 |
| $\omega$ (1420) | $0^- (1^{--})$ | u/d | $\omega$ (782)$U_2$ | $\rho$ (770) $U_2e_2$ | 1414.3 | 1419.0 | -4.7 |
| $\omega$ (1650) | $0^- (1^{--})$ | u/d | $\omega$ (1420)$Ue_2$ | $\rho$ (770) $UXm_2$ $e_2$ | 1649.9 | 1649.0 | 0.9 |
| <u>u and d in major decay modes, s in minor decay modes</u> | | | | | | | |
| $f_0$ (980) | $0^+ (0^{++})$ | u/d/s | $\pi^0$ $X_2$ | $\pi°X_2$ | 980.3 | 980.0 | 0.3 |
| $a_O$ (980) | $1^- (0^{++})$ | u/d/s | $f_0$ (980)$e_2$ | $\eta$ (547)X | 982.6 | 984.0 | 4.4 |
| $b_1$(1235) | $1^+ (1^{+-})$ | u/d/s | u + d + s ≈ $U_3m_4e_2$ | $\pi^\pm$ $U_2m_2e$ | 1228.8 | 1229.5 | -0.7 |
| $a_1$ (1260) | $1^- (1^{++})$ | u/d/s | 2MM0 + MM2 ≈ $\eta$'(958) $m_4$ | $\pi°U_3m_2e_2$ | 1228.8 | 1230.0 | -1.2 |
| $f_2$ (1270) | $0^+ (2^{++})$ | u/d/s | $f_0$ (980)$m_4$ | $\pi°U_2Xme_2$ | 1273.3 | 1275.4 | -2.1 |
| $f_1$ (1285) | $0^+ (1^{++})$ | u/d/s | $f_2$ (1270) $e_2$ | $\pi°U_3Xm_3$ | 1289.6 | 1281.9 | 7.7 |
| $a_2$ (1320) | $1^- (2^{++})$ | u/d/s | $a_O$ (980)U | $\rho$ (770)$Um_3$ | 1318.0 | 1318.0 | -1.0 |
| $\rho$ (1450) | $1^+ (1^{--})$ | u/d/s | $a_2$ (1320)$m_2$ | $\rho$ (770)$Um_5e_2$ | 1465.0 | 1465.0 | 0.0 |
| $a_0$ (1450) | $1^- (0^{++})$ | u/d/s | $a_0$ (980)Xm | $\eta$ (547)$X_2m$ | 1472.7 | 1474.0 | -1.3 |
| $f_0$ (1500) | $0^+ (0^{++})$ | u/d/s | $f_1$ (1285)$m_3e_2$ | $\pi°U_3X$ | 1507.7 | 1507.0 | 0.7 |
| $\omega_3$ (1670) | $0^- (3^{--})$ | u/d/s | 4MM1≈ $U_4$ X | $\pi°U_4m_4$ | 1664.0 | 1667.0 | -3.0 |
| $\pi_2$ (1670) | $1^- (2^{-+})$ | u/d/s | $U_4Xe_2$ | $\pi°U_4m_4$ $e_2$ | 1672.0 | 1670.0 | 2.0 |
| $\rho_3$ (1690) | $1^+ (3^{--})$ | u/d/s | s + s + s | $\rho$ (1450)$m_3e_2$ | 1683.1 | 1691.0 | -7.9 |
| $\rho$ (1700) | $1^+ (1^{--})$ | u/d/s | $\rho_3$ (1690)$e_2$ | $\pi^\pm U_5e_2$ | 1700.9 | 1700.0 | 0.9 |
| $\pi$(1800) | $1^- (0^{-+})$ | u/d/s | $a_0$ (1450)U | $\pi^\pm U_4Xe2$ | 1808.7 | 1801.0 | 7.7 |
| $a_4$(2040) | $1^- (4^{++})$ | u/d/s | $\omega_3$ (1670)$Ue_2$ | $\pi^\pm U_6e2$ | 2005.3 | 2011.0 | -5.7 |
| $f_4$ (2050) | $0^+$ (4++) | u/d/s | $a_4$(2040)$e_2$ | $\pi°U_5m_5$ | 2043.1 | 2025.0 | 9.1 |
| <u>s in major decay modes</u> | | | | | | | |
| $\Phi$ (1020) | $0^- (1^{--})$ | s/u/d | $U_3me_2$ | $U_3me_2$ | 1017.6 | 1019.4 | -1.8 |
| $\eta$ (1405) | $0^+ (0^{-+})$ | s/u/d | $\Phi$ (1020)Um | K°(498)$U_2m_4$ | 1409.0 | 1410.3 | -1.3 |
| $f_1$ (1420) | $0^+ (1^{++})$ | s/u/d | $U_3Xm$ | K°(498)$U_2m_4e_2$ | 1421.0 | 1426.3 | -5.3 |
| $\eta$ (1475) | $0^+ (0^{-+})$ | s/u/d | $\eta$ (1405)m | K°(498)$U_2m_5$ | 1483.0 | 1476.0 | 7.0 |
| $f_2'$ (1525) | $0^+ (2^{++})$ | s/u/d | $m_4U_4$ | K°(498)$UXm_4$ | 1528.7 | 1525.0 | 3.7 |
| $f_0$ (1710) | $0^+ (0^{++})$ | s/u/d | $f_1$ (1420)$m_4e_2$ | K*(892)$U_2m_2e_2$ | 1680.5 | 1710.0 | 0.5 |
| $\Phi$ (1680) | $0^- (1^{--})$ | s/u/d | f '$_2$(1525)$m_2e_2$ | K°(498)$U_3m_4$ | 1717.3 | 1680.0 | 3.3 |
| $\Phi_3$ (1850) | $0^- (3^{--})$ | s/u/d | $U_6$ | K°(498)$U_3X$ | 1857.4 | 1854.0 | 3.4 |
| $f_2$ (2010) | $0^+ (2^{++})$ | s/u/d | $\Phi_3$ (1850)$m_2e_2$ | O(1020)$U_2m_5e_2$ | 2012.7 | 2011.0 | 1.7 |
| $f_2$ (2300) | $0^+ (2^{++})$ | s/u/d | $f_2'$ (1525)$m_2U_2$ | O(1020)$U_2Xm_3e_2$ | 2292.8 | 2297.0 | -4.2 |
| $f_2$ (2340) | $0^+ (2^{++})$ | s/u/d | $f_2'$ (2010)U | O(1020)$U_4me_2$ | 2341.4 | 2339.0 | 2.4 |

## **Table 12.** Strange, Charmed, and Bottom Mesons

| Meson | $J^{pc}$ | Quark | Initial formula and sequence | Decay formula | Calculated mass. | Observed mass | Difference |
|---|---|---|---|---|---|---|---|
| Light strange mesons | | | | | | | |
| $K^{\pm}$ (494) | $0^-$ | $u\bar{s}$, $\bar{u}s$ | MM1me = Xme | Xme | 494.8 | 493.7 | 1.1 |
| K° (498) | $0^-$ | $d\bar{s}$ | $MM1 + m + e_2 \approx Xme_2$ | $Xme_2$ | 498.2 | 497.7 | 0.5 |
| $K^*_0$ (1430) | $0^+$ | $d\bar{s}$ | $K°(498)m_4U_2$ | $K°(498)\ m_4U_2$ | 1413.0 | 1412.0 | 1.0 |
| $K^*\ (892)^{\pm}$ | $1^-$ | $u\bar{s}$, $\bar{u}s$ | $u + s \approx K^{\pm}$ (494)Um | $K^{\pm}$ (494)mU | 894.5 | 891.6 | 2.9 |
| K* (892)° | $1^-$ | $d\bar{s}$ | $d + s \approx K^*\ (892)^{\pm}\ e$ | $K^*\ (892)^{\pm}e$ | 896.2 | 896.1 | 0.1 |
| $K_1$ (1270) | $1^+$ | $d\bar{s}$ | $K^0(498)m2U_2$ | $K^0(498)U_2\ m_2$ | 1272.9 | 1273.0 | -0.1 |
| $K_1$ (1400) | $1^+$ | $d\bar{s}$ | $K_1\ (1270)m_2$ | $K^*\ (892)°Xme_2$ | 1394.3 | 1402.0 | -7.7 |
| K* (1410) | $1^-$ | $d\bar{s}$ | $K°(498)m_4\ U_2$ | $K°(498)\ U_2m_4$ | 1413.0 | 1414.0 | -1.0 |
| $K^*_2(1430)^{\pm}$ | $2^+$ | $u\bar{s}$, $\bar{u}s$ | $K_1\ (1270)m_2\ e$ | $K^0(498)U_2m_4e_3$ | 1424.4 | 1425.6 | -1.2 |
| $K^*\ _2(1430)^0$ | $2^+$ | $d\bar{s}$ | $K^*_2(1430)^{\pm}\ e$ | $K^0(498)U_3$ | 1437.2 | 1432.4 | -4.6 |
| K* (1680) | $1^-$ | $d\bar{s}$ | $K°(498)m_4U_3$ | $K^0(498)U_3m_4$ | 1717.3 | 1717.0 | 0.3 |
| $K_2$ (1770) | $2^-$ | $d\bar{s}$ | $K^*_2\ ((1430)^0\ Ue_2$ | $K^*\ _2(1430)^0Ue_2$ | 1771.2 | 1773.0 | -1.8 |
| $K_3$* (1780) | $3^-$ | $d\bar{s}$ | $K_2\ (1770)\ e_2$ | $K^0(498)U_2Xm_3e_2$ | 1771.1 | 1776.0 | -4.9 |
| $K_2$ (1820) | $2^-$ | $d\bar{s}$ | $K_1(1400)X$ | $K^0(498)U_4me_2$ | 1816.0 | 1816.0 | 3.7 |
| $K_4$* (2045) | $4^+$ | $d\bar{s}$ | $K^*_3(1780)\ m_4$ | $K^0(498)U_5$ | 2046.0 | 2045.0 | 1.0 |
| Charmed strange mesons | | | | | | | |
| $D_s^{\pm}$ | $0^-$ | $s\bar{c}$, $\bar{s}c$ | $K_2\ (1820)m_2\ e$ | $K^0(498)U_4m_3e_2$ | 1968.3 | 1968.5 | -8.6 |
| $D^*{}_s^{\pm}$ | $1^-$ | $s\bar{c}$, $\bar{s}c$ | $D_s^{\pm}\ m_2$ | $D_s^{\pm}\ m_2$ | 2108.4 | 2112.1 | -3.7 |
| $D_{s1}(2536)^{\pm}$ | $1^+$ | $s\bar{c}$, $\bar{s}c$ | $D^*{}_s^{\pm}$ X | D* (2007)$Um_3$ | 2535.4 | 2535.4 | 12.1 |
| $D_{sJ}(2573)^{\pm}$ | $2^+$ | $s\bar{c}$, $\bar{s}c$ | $D^*{}_s^{\pm}m_2$ U | $D^0U_2m_3$ | 2569.7 | 2572.4 | -2.7 |
| Charmed mesons | | | | | | | |
| D(1865)° | $0^-$ | $u\bar{c}$ | $MM3 \approx\ U_6\ e_2$ | $U_6\ e_2$ | 1860.7 | 1864.5 | -3.8 |
| $D(1869)^{\pm}$ | $0^-$ | $d\bar{c}$, $\bar{d}c$ | D(1864)°e | $U_6\ e_3$ | 1864.1 | 1869.3 | -5.2 |
| $D^*\ (2007)^o$ | $1^-$ | $u\bar{c}$ | $u + c \approx D°(1864)°\ m_2$ | $D^0\ m_2$ | 2004.6 | 2006.7 | -2.1 |
| $D^*(2010)^{\pm}$ | $1^-$ | $d\bar{c}$, $\bar{d}c$ | $d + c \approx D^*\ (2007)^o\ e$ | $D^0\ m_2e$ | 20009.1 | 2010 | -0.9 |
| $D_1$ (2420)° | $1^+$ | $u\bar{c}$ | $D^*(2010)^{\pm}$ X | $D^*(2010)^{\pm}Ue$ | 2415.4 | 2422.2 | -6.8 |
| $D_2$* (2460)° | $2^+$ | $u\bar{c}$ | $U_8$ | $U_8$ | 2461.5 | 2458.9 | 2.6 |
| $D_2*(2460)^{\pm}$ | $2^+$ | $d\bar{c}$, $\bar{d}c$ | $U_8\ \ e$ | $U_8e$ | 2466.1 | 2459 | 7.1 |
| Bottom mesons | | | | | | | |
| $B^{\pm}$ | $0^-$ | $u\bar{b}$, $\bar{u}b$ | $MM5Um \approx mU_{17}\ e_2$ | $mU_{17}\ e_2$ | 5279.0 | 5279.0 | 0.0 |
| B° | $0^-$ | $d\bar{b}$ | $B^{\pm}e$ | $mU_{17}\ e_3$ | 5282.4 | 5279.4 | 3.0 |
| B* | $1^-$ | $d\bar{b}$ | $MM5\ X \approx XU_{16}\ e_2$ | $B^0\ e_{12}$ $XU_{16}\ e_2$ c | 5324.7 | 5325 | -0.3 |
| Bs | $0^-$ | $s\bar{b}$ | B* $me_2$ | $D_s^{\pm}\ U_9Xm_3$ | 5364.4 | 5369.6 | -5.2 |

**Table 13:** c $\bar{c}$ mesons and b $\bar{b}$ mesons

| Meson | $J^{pc}$ | Quark | Initial formula and sequence | Decay formula | Calculated mass. | Observed mass | Difference |
|---|---|---|---|---|---|---|---|
| c $\bar{c}$ mesons | | | | | | | |
| $\eta_c$ (1s) | $0^{-+}$ | c $\bar{c}$ | MM4 ≈ $U_9$ | $U_9$ | 2976.9 | 2979.7 | -2.8 |
| J/$\psi$ | $1^{--}$ | c $\bar{c}$ | MM4 ≈ $N_2U_4$ | $N_2U_4$ | 3096.7 | 3096.9 | -0.2 |
| $\chi_c$ (1p) | $0^{++}$ | c $\bar{c}$ | c + c ≈ $\eta_c$ (1s)X $e_2$ | $\pi^0$ $U_9m_4e_3$ | 3421.0 | 3415.1 | 5.9 |
| $\chi_{c1}$ (1p) | $1^{++}$ | c $\bar{c}$ | J/$\psi$ X | J/$\psi$X | 3517.1 | 3510.5 | 6.6 |
| $\chi_{c2}$ (1p) | $2^{++}$ | c $\bar{c}$ | $\chi_{c1}$ (1p) $m_2$ | J/$\psi$X$e_{10}$ | 3552.2 | 3556.2 | -4.0 |
| $\psi_{2s}$ | $1^{--}$ | c $\bar{c}$ | $\eta_c$ (1s)m $_2$X $_2e_2$ | J/$\psi$X$m_2e_8$ | 3554.7 | 3686.0 | -1.5 |
| $\psi$ (3770) | $1^{--}$ | c $\bar{c}$ | $\chi_{c2}$ (1p)$m_3$ | $D^0X_4m_3e_2$ | 3763.2 | 3769.9 | -6.7 |
| $\psi$ (4040) | $1^{--}$ | c $\bar{c}$ | J/$\psi$ N | $D^0X_5me_2$ | 4043.3 | 4040.0 | 3.3 |
| $\psi$ (4160) | $1^{--}$ | c $\bar{c}$ | $\psi_{2s}$mU | $\pi^0N_4Um_2$ | 4158.0 | 4159.0 | -1.0 |
| $\psi$ (4415) | $1^{--}$ | c $\bar{c}$ | $\chi_c$ (1p)$U_3e_2$ | $\pi^0N_4U_2$m | 4418.8 | 4415.0 | 3.8 |
| b $\bar{b}$ mesons | | | | | | | |
| $\Upsilon$ (1s) | $1^{--}$ | b $\bar{b}$ | $N_7$ $U_6$ $X_3$ $e_2$ | n'(958)U25m3e2 | 9460.2 | 9460.3 | -0.1 |
| $\chi_{b0}$ (1p) | $0^{++}$ | b $\bar{b}$ | $N_{10}U_3$ | $\Upsilon$ (1s)$m_5e_{14}$ | 9859.2 | 9859.9 | -0.7 |
| $\chi_{b1}$ (1p) | $1^{++}$ | b $\bar{b}$ | $\Upsilon$ (1s)X$e_2$ | $\Upsilon$ (1s) X$e_2$ | 9888.5 | 9892.7 | -4..2 |
| $\chi_{b2}$ (1p) | $2^{++}$ | b $\bar{b}$ | $\chi_{b1}$ (1p)$e_2$ | $\Upsilon$ (1s)X$e_8$ | 9908.8 | 9912.6 | -3.8 |
| $\Upsilon$ (2s) | $1^{--}$ | b $\bar{b}$ | $\Upsilon$ (1s)X$m_2$ | $\Upsilon$ (1s)X$m_2$ | 10020.5 | 10023.3 | -2.8 |
| $\chi_{b0}$ (2p) | $0^{++}$ | b $\bar{b}$ | $\Upsilon$ (2s)$m_3$ | $\Upsilon$ (2s)$m_3$ | 10233.4 | 10232.1 | -1.3 |
| $\chi_{b1}$ (2p) | $1^{++}$ | b $\bar{b}$ | $\Upsilon$ (2s)$m_2$ $e_2$ ≈ $U_{31}$ | $\Upsilon$ (2s)$m_2$ $e_2$ | 10241.4 | 10255.2 | -13.8 |
| $\chi_{b2}$ (2p) | $2^{++}$ | b $\bar{b}$ | $\chi_{b1}$ (2p) $e_2$ | $\Upsilon$ (2s)$m_3e_{10}$ | 10268.6 | 10268.5 | 0.1 |
| $\Upsilon$ (3s) | $1^{--}$ | b $\bar{b}$ | $\Upsilon$ (2s) U | $\Upsilon$ (2s) + U | 10354.1 | 10355.2 | -1.1 |
| $\Upsilon$ (4s) | $1^{--}$ | b $\bar{b}$ | b + b = MM6 ≈ $U_{32}$ | B°U16e2 | 10579.7 | 10580.0 | -0.3 |
| $\Upsilon$ (10860) | $1^{--}$ | b $\bar{b}$ | $\Upsilon$ (4s) $m_4$ | U32m4 | 10864.7 | 10865 | -0.3 |
| $\Upsilon$ (11020) | $1^{--}$ | b $\bar{b}$ | $\Upsilon$ (3s)$U_2$ | U32Xe2 | 11012.8 | 11019 | -6.2 |

## 6. *QCD and the three sets of seven orbitals*

The reasons for the confinement of quarks in colorless hadrons, asymptotic freedom, and colorflavor locking in QCD [14] can be found in the three sets of seven orbitals. Essentially, auxiliary dimensional orbital explains the confinement of quarks, principal dimensional orbital accounts for asymptotic freedom, and hadronic dimensional orbital is responsible for colorflavor locking.

The confinement of individual quarks in colorless hadrons implies hidden individual quarks. Such hidden individual quark is the result of hidden auxiliary dimensional orbital [7], which is a part of individual quark. The hidden individual quark manifests in the observation that only particle with integral electric charge and hypercharge can exist in isolation, and individual quark with fractional electric charge and hypercharge cannot exist in isolation. Therefore, quarks can exist only as the quark composites with integral electric charges and hypercharges. The composites of quarks are hadrons, including mesons and baryons. $B_6$ ($\pi_{1/2}$) is the gauge boson to maintain integral electric charges and hypercharges in hadrons. This gauge boson becomes eight gluons with SU(3) that change the color charges from quarks in order to confine quarks in colorless hadrons.

Asymptotic freedom in QCD states that the coupling strength of gluons decreases with increasing temperature. In terms of principal dimensional orbital and auxiliary dimensional orbital, the increasing temperature causes the occupation of quark to shift toward simpler principal dimensional orbital. All quarks include d = 7 (Fig. 1) as principal dimensional orbital, so at increasing temperature, quarks increasingly occupy d = 7 principal orbital. In d = 7, the lepton is $\nu_\mu$, which has chiral symmetry, so at very high temperature, quark in d = 7 has chiral symmetry, which transforms high-mass constituent quarks into low-mass current quarks. Being in principal dimensional orbital, quarks are not hidden, and do not need to couple with gluons. Thus, at very high temperature, quarks in principal dimensional orbital are free quasiparticles with little or no coupling with gluons as shown in the jet-like appearance for the production of hadrons in electron-positron annihilations at high energy.

At low temperature, QCD shows "color superconductivity", resulting in colorflavor locking. In colorflavor locking, flavors and colors are correlated. The gluons become massive and electrically charged. The quark charges are shifted. The electric charges of these particles all become integral multiples of the charge of electron. The reason for colorflavor locking is that at decreasing temperature, the occupation of quarks shifts increasingly toward more complicate hadronic dimensional orbital. In hadronic dimensional orbital, $B_6$ ($\pi_{1/2}$) as gluons become massive, acquires $\pm$ electric charge, and becomes a part of pion and other hadrons as shown in the constituent quark model. Essentially, at low temperature, QCD becomes CQD (constituent quark dynamics).

In summary, quarks start with the occupation of both principal dimensional orbital and hidden auxiliary dimensional orbital, resulting in the confinement of quarks. At increasing temperature, the occupation of quarks increasing shifts toward simpler principal dimensional orbital, resulting in asymptotic freedom. At decreasing temperature, the occupation of quarks shifts increasingly toward more complicate hadronic dimensional orbital, resulting in colorflavor locking.

## *7. Conclusion*

All elementary particles and hadrons can be explained by the three sets of seven orbitals: principal dimensional orbital (Fig.1), auxiliary dimensional orbital (Fig.1), and hadronic dimensional orbital (Fig.2). Principal dimensional orbital derived from space-time dimension, varying speed of light, and varying supersymmetry explains gauge bosons and low-mass leptons. Auxiliary dimensional orbital derived from principal dimensional orbital accounts for heavy leptons and individual quarks. Hadronic dimensional orbital derived from auxiliary dimensional orbital is responsible for hadrons, the composites of individual quarks.

As the periodic table of elements is derived from atomic orbital, the periodic table of elementary particles (Fig. 1 and Table 2) is derived from principal dimensional orbital and auxiliary dimensional orbital. All leptons, quarks, and gauge bosons can be placed in the periodic table. The calculated masses for gauge bosons, leptons. and quarks are listed in Table 1 and Table 3. The calculation involves the extended Barut mass formula.

Hadronic dimensional orbital generates hadronic marks, representing stable hadrons. Hadronic dimensional orbital relates to the Polazzi mass formula for stable hadrons. Quarks from auxiliary dimensional orbital and hadronic marks from hadronic dimensional orbital constitute the quark-mark formulas (Tables 7 and 10) for hadrons. The transitions among quarks are through mass blocks consisting of various basic fermions, such as $\pi_{1/2}$, u quark, e, p, and n. The compositions from mass blocks are the mass block formulas. The mass block formula corresponds to MacGregor-Aker constituent quark model. The quark-mark formula is matched by the mass block formula, and the mass block formula constitutes the precise composition for the mass of hadron. The mass block fermions and the calculated masses are listed for baryons in Tables 8 and 9, and for mesons are listed in Tables 11, 12, and 13. QCD, essentially, describes the different occupations of quarks in the three sets of seven orbitals at different temperatures.

The masses of elementary particles can be calculated using only four known constants: the number of the extra spatial dimensions in the eleven dimensional membrane, the mass of electron, the mass of Z°, and $\alpha_e$. The calculated masses for elementary particles and hadrons are in good agreement with the observed masses. For examples, the calculated masses for the top quark, neutron, and pion are 176.5 GeV, 939.54MeV, and 135.01MeV in excellent agreement with the observed masses, 174.3 ± 5.1GeV, 939.57 MeV, and 134.98 MeV, respectively.

## References

* chung@wayne.edu P.O. Box 180661, Utica, Michigan 48318, USA